\documentclass[preprint2,10pt,column]{aastex}

\usepackage[
  a4paper, mag=1000, includefoot,
  left=2.0cm, right=1cm, top=2cm, bottom=2cm, headsep=1cm, footskip=1cm
]{geometry}
\usepackage[T2A]{fontenc}
\usepackage[utf8]{inputenc}
\usepackage[english]{babel}
\usepackage{babelbib}
\usepackage{rotating}
\usepackage{floatrow}
\usepackage{longtable}
\usepackage{multirow}
\usepackage{lscape}
\usepackage{array}
\usepackage{cleveref}
\usepackage{graphicx}

\ifpdf\usepackage{epstopdf}\fi
\usepackage{journals}

\def\apjl{Astrophys. J. Let}
\def\apjs{Astrophys. J. Suppl}
\def\aap{Astron. and Astrophys}

\def\gee{ \, \lower 1mm\hbox{$\,{\buildrel > \over{\scriptstyle\scriptstyle\sim} }\displaystyle \,$}}
\def\lee{ \, \lower 1mm\hbox{$\,{\buildrel < \over{\scriptstyle\scriptstyle\sim} }\displaystyle \,$}}
\def\ltsima{$\; \buildrel < \over \sim \;$}
\def\simlt{\lower.5ex\hbox{\ltsima}}
\def\gtsima{$\; \buildrel > \over \sim \;$}
\def\simgt{\lower.5ex\hbox{\gtsima}}

\def\H2{H$_{2}$}

\def\roH2{$\rho_{\mbox{\footnotesize \H2}}$}

\def\MH2{M$_{\textrm{H}_2}$}

\def\mgfe{[Mg/Fe]\,}
\def\Ha{H$\alpha$} 
\def\Hb{H$\beta$} 
\def\i{\,{\small I}}
 
\def\iii{\,{\small III}}
\def\nb{\textsc{NBursts}}

\shorttitle{Isolated Lenticular galaxies}
\shortauthors{I.Yu.~Katkov}

\begin{document}

\title{\textnormal{The translate of strongly abridged PhD thesis:}\\
\MakeUppercase{Isolated lenticular galaxies: properties and evolution}\footnote{The PhD thesis submitted to Lomonosov Moscow State University, Sternberg Astronomical Institute. PhD supervisor: Prof. Olga K. Sil'chenko (SAI MSU).}}
\author{Ivan Yu. Katkov}
\affil{Lomonosov Moscow State University \\ 
Faculty of Physics, Sternberg Astronomical Institute \\ 
Universitetsky pr., 13, Moscow, 119991, Russia}
\email{katkov.ivan@gmail.com} 

\begin{abstract}
This work is dedicated to investigation of galaxies that do not fit into a common scenario of galaxy
formation -- isolated lenticular galaxies. We have studied stellar populations and ionized gas content of
a sample of 22 lenticular galaxies (among those 4 targets have appeared to be of erroneous morphological classification) by undertaking deep long-slit spectroscopy with the Russian 6-m telescope and with the Southern African
Large Telescope (SALT). The obtained average ages of the stellar populations in bulges and discs covers a
wide range between 1.5 and $>15$ Gyr, that indicates the absence of distinct epoch of their stellar content formation.
In contrast to galaxies in groups and clusters, the stellar population ages in bulges and discs of isolated
lenticulars tend to be equal, that supports the inefficiency of the bulge rejuvenation in sparse environment. Almost all the lenses and rings possess intermediate ages of the stellar populations, within the range of $2-5$ Gyr. By analyzing the emission-line spectra of galaxies, we have found that 13 out of 18 ($72\pm11$ \%) objects of our sample possess extended emission-line structures; among those, 6 galaxies ($46\pm14$ \%) demonstrate decoupled gas kinematics with respect to their stellar discs. We have found starforming off-nuclear regions in 10 galaxies; their gas oxygen abundances are nearly solar that implies tidal gas accretion from gas-rich dwarf satellites rather than accretion from cosmological large-scale structure filaments.

\end{abstract}

\tableofcontents

\section{Introduction}

One of the central topics in current extragalactic astronomy is how galaxies form and how their properties change through cosmic times. This is a particularly difficult task if concerning lenticular galaxies since this class of objects show a diversity in their properties that goes beyond present day simulations. 

The standard scenario of formation of lenticular galaxies relates to transformation of spiral galaxies into lenticulars by dense environment effect: ram pressure in hot intracluster/intragroup medium \citep{gunn_gott_1972,quilis_2000}, gravitational tides and harassment \citep{byrd_1990,moore_1996}, direct encounters of galaxies  \citep{spitzer_1951, icke_1985}, starvation \citep{larson_1980}. However 15\% of nearby field galaxies are lenticulars \cite{naim_1995}, and there are examples of strictly isolated S0 \cite{sulentic_2006}. The study of isolated lenticulars by using deep optical spectroscopy methods provided information not only for central part of galaxies but also for disc components, is important issue because it provides a crucial point for the testing of scenario  formation of lenticulars at all.

\section{Sample of galaxies}

We have compiled our sample of isolated S0 galaxies basing on the approaches recently developed by the team of Igor Karachentsev, Dmitry Makarov, and co-authors (Special Astrophysical Observatory, Russia). Their group-finding algorithm that takes into account individual characteristics of galaxies has been already used to study the properties of isolated galaxies \citep{karachentsev_isol_2011}, pair \citep{karachentsev_pairs_2008}, triple \citep{makarov_triplets_2009} systems of galaxies and galaxy groups \citep{makarov_groups_2011}. The LOG catalogue (``Local Orphan Galaxies'') of 520 extremely isolated galaxies has been presented in \citet{karachentsev_isol_2011}. However, LOG catalogue contains only 17 early type galaxies ($T<0$). Thus, we composed list of 281 isolated S0 galaxies by slightly weakened the isolation criteria based on their approach. Using that list of galaxies we chosen targets for spectral observation. The details about selection criteria will be presented in \citet{katkov_bull_stpop} and in the prepared paper with SALT data results \citep{katkov_salt}.

\section{Observations and data analysis}
The observations of galaxies were carried out with 6-m Russian telescope operated by Special Astrophysical Observatory at the Russian Academy of Sciences (Nizhnij Arkhyz,Russia) as well as with Southern African Large Telescope (Sutherland, SA) during period 2011-2013 years under observation programs dedicated to studies of isolated S0s (P.I.: Olga~K.~Sil'chenko (SAI MSU), co-Is: Alexey~Yu.~Kniazev (SAAO, SAI MSU) and Ivan~Yu.~Katkov). We observed 22 targets, but subsequent analyses of Kormendy diagram reveals that 4 targets among 22 objects is ellipticals. So, the analysis of stellar populations and ionized gas carried out based only 18 galaxies.

\subsection{6-m telescope BTA}
The spectral data for most galaxies were obtained using the new focal reducer SCORPIO-2 \citep{scorpio2} maintained at the prime focus of the telescope. The grism VPHG1200@540 provides spectral resolution $\approx$ 4~\AA\ in spectral range $3800-7300$ \AA. 

This spectral range included a set of strong absorption and emission features, making it suitable to study both internal stellar and gaseous kinematics and the stellar populations of the galaxy. The $1''$-width slit was oriented along major axes of galaxies and was $6'$ in length, which provided the possibility of using the edge spectra to evaluate the sky background. The CCD chip E2V CCD42-90, with a format of 2048x4600, used in the 1x2 binning mode provided a spatial scale of $0''.357$ pixel$^{-1}$ and a spectral sampling of 0.84 \AA\ pixel$^{-1}$.

The preliminary data reduction was identical to that applied to the lenticular galaxy NGC~7743 described in \citet{n7743}. Briefly, the primary data reduction was comprised of bias subtraction, flat-fielding, cosmic ray hit removal, and building the wavelength solution using the He-Ne-Ar arc-line spectra. To subtract the sky background, we invented a rather sophisticated approach.We constructed the spectral line-spread function (LSF) model varied along and across the wavelength direction by using the twilight spectrum \citep{skysubtraction}. The final stages of the long-slit spectra reduction were night sky spectrum subtraction taking into account the LSF variations, linearization, and accounting for spectral sensitivity variation using the spectrum of a spectrophotometrical standard star. The error frames were computed using the photon statistics and

\subsection{SALT telescope}

The observations were conducted with the Robert Stobie Spectrograph
\citep{Burgh_2003_RSS,Kobulnicky_2003_rss} at Southern African Large Telescope. 
The long-slit spectroscopy mode of the RSS was used
with a 1.25 arcsec slit width for the most of observations.  The total time of
one observational block with SALT is limited by the track-time of about an hour
for our targets.  For this reason and because SALT is a queue-scheduled
telescope, most of our galaxies were observed more than once and all
observations were done during differents nights. The total time exposure per object between 0.5 and 3 hours.  The slit was oriented along the major axis
for all galaxies. The grating GR900 was used for our program to cover finally
the spectral range of $3760-6860$ \AA\ with a final reciprocal dispersion of
$\approx0.97$ \AA\ pixel$^{-1}$ and FWHM spectral resolution of $4.7-4.8$ \AA.  The seeing during observations was in the range
$1.5-3.0$ arcsec.  The RSS pixel scale is $0''.129$ and the
effective field of view is 8\arcmin\ in diameter.  We utilised a binning
factors of 2 and 4 to give a final spatial sampling of $0''.258$ pixel$^{-1}$
and $0''.516$ pixel$^{-1}$ respectively.  Spectrum of an Ar comparison arc was
obtained to calibrate the wavelength scale after each observation as well as
spectral flats were observed regularly to correct for the pixel-to-pixel
variations. Spectrophotometric standard stars were observed during twilights,
after observations of objects for the relative flux calibration.

Primary data reduction was done with the SALT science pipeline \citep{Crawford_2010_SALTpipeline}.
After that, the bias and gain corrected and mosaicked long-slit data were
reduced in the way described in \citet{Kniazev_2008}.  The accuracy of the spectral
linearisation was checked using the sky line [O\i]~$\lambda$5577; the RMS
scatter of its wavelength measured along the slit is 0.04~\AA.  The slit length
is approximately 8\arcmin, so sky spectra from the slit edges were used to
estimate the background during the galaxy exposures.

\subsection{Extraction of kinematics and stellar population properties}

To derive information about stellar and ionized gas kinematics we
first fitted the stellar absorption spectra by the PEGASE.HR high resolution stellar population models \citet{pegasehr} convolved with a parametric line-of-sight (LOS) velocity distribution by applying \nb\ full spectral fitting technique \citet{nbursts_a,nbursts_b}. Before the minimization procedure, the model grid of stellar population spectra is convolved with the LSF. Multiplicative Legendre polynomials are also included to take into account possible internal dust reddening and residual spectrum slope variations due to the errors in the assumed instrument spectral response. Ionized-gas emission lines and remnants of the subtracted strong airglow lines do not affect the solution due to masking the narrow 15 \AA -wide regions around them. The resulted stellar parameters are LOS velocity $v$, velocity dispersion $\sigma$, higher order Gauss–Hermite coefficient $h_3$,$h_4$ and the stellar population parameters: the age $T$ and metallicity [Z/H]. 
The radial profiles of the parameters for SCORPIO data are partly shown in \citet{ilg_gas}, also will be appeared in \citet{katkov_bull_stpop}; the paper with SALT data results is in preparation \citep{katkov_salt}.

For the full spectral fitting we utilized solar alpha-enhanced stellar population models which doesn't provide properties of alpha-enrichments of stellar population. In order to estimate [Mg/Fe] also we measured Lick indices and estimated [Mg/Fe] by comparison with the evolutionary SSP models by \citep{Thomasstpop}.

\section{Stellar populations}
\begin{figure*}
\centerline{
\includegraphics[width=0.45\textwidth]{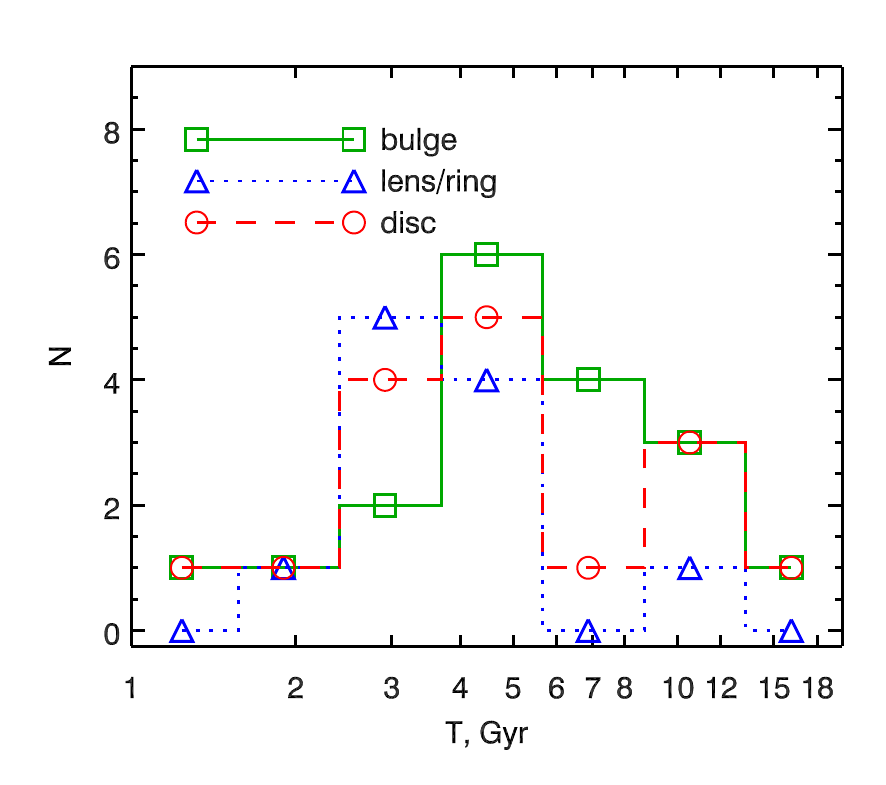}
\includegraphics[width=0.45\textwidth]{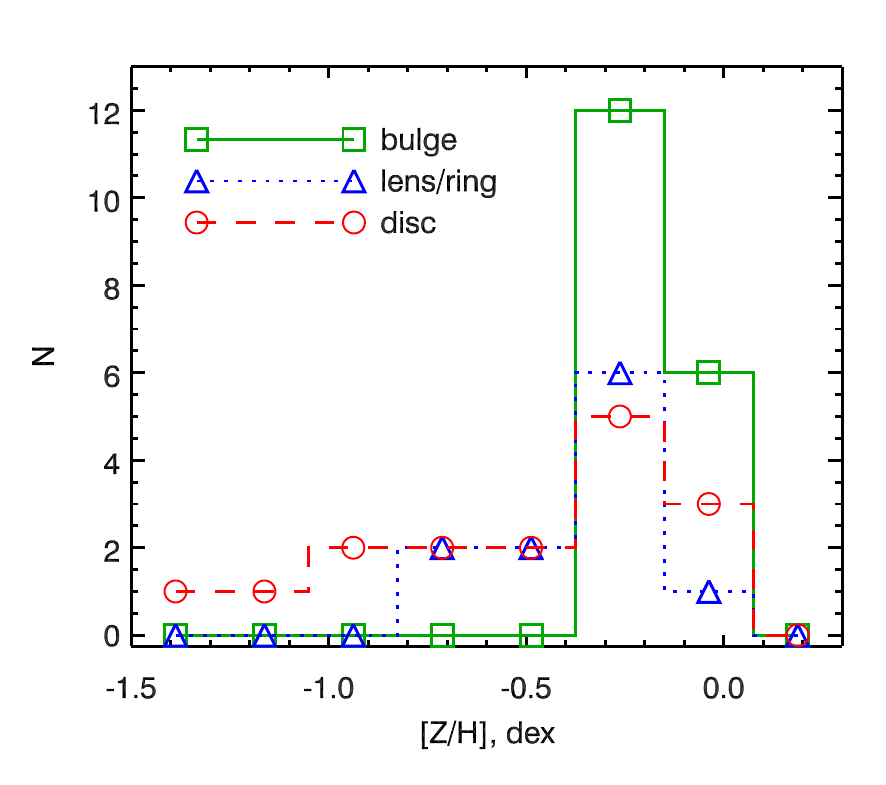}
}
\centerline{
\includegraphics[width=0.45\textwidth]{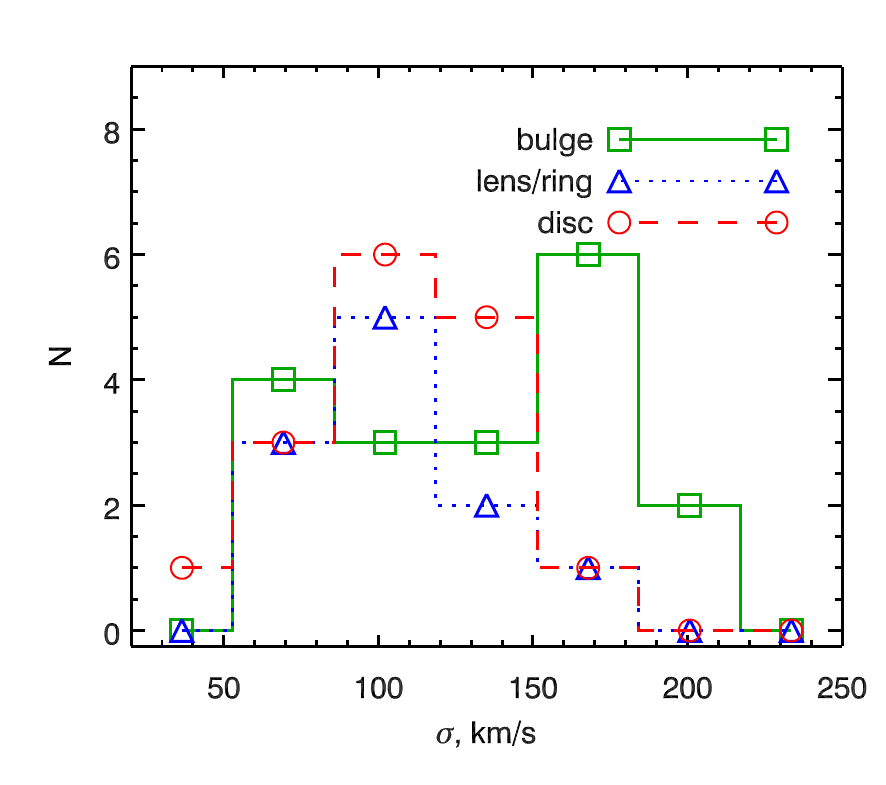}
\includegraphics[width=0.45\textwidth]{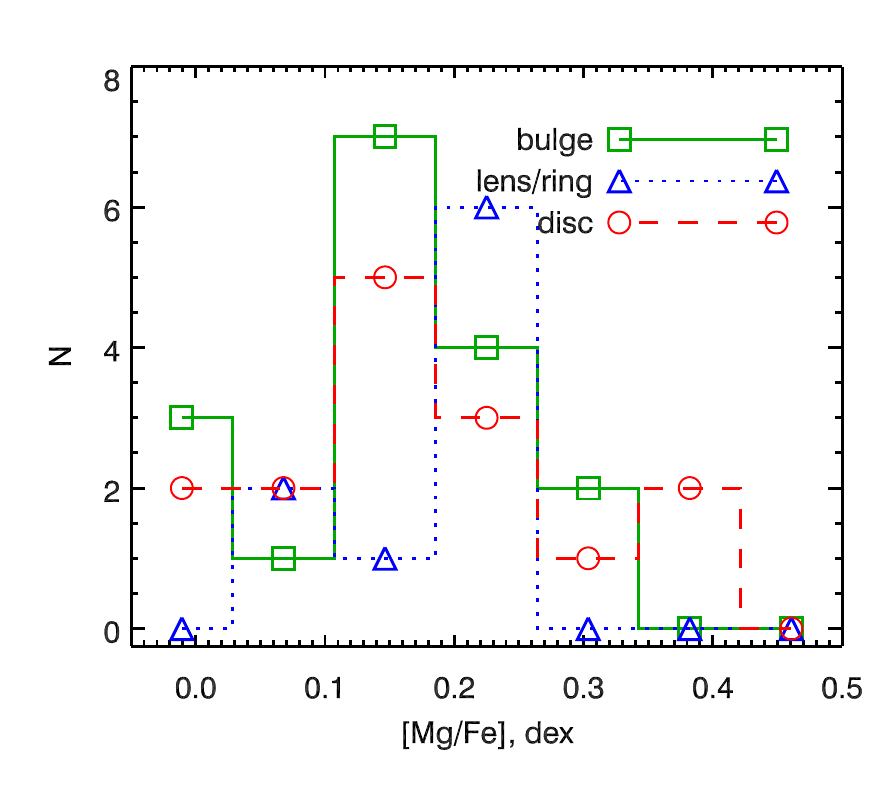}
}
\caption{Histogram for ages metallicities,velocity dispersion and \mgfe for different stellar substructures of galaxies.}\label{pics_stpop_hist}
\end{figure*}

\begin{figure*}
\centerline{
\includegraphics[width=0.3\textwidth]{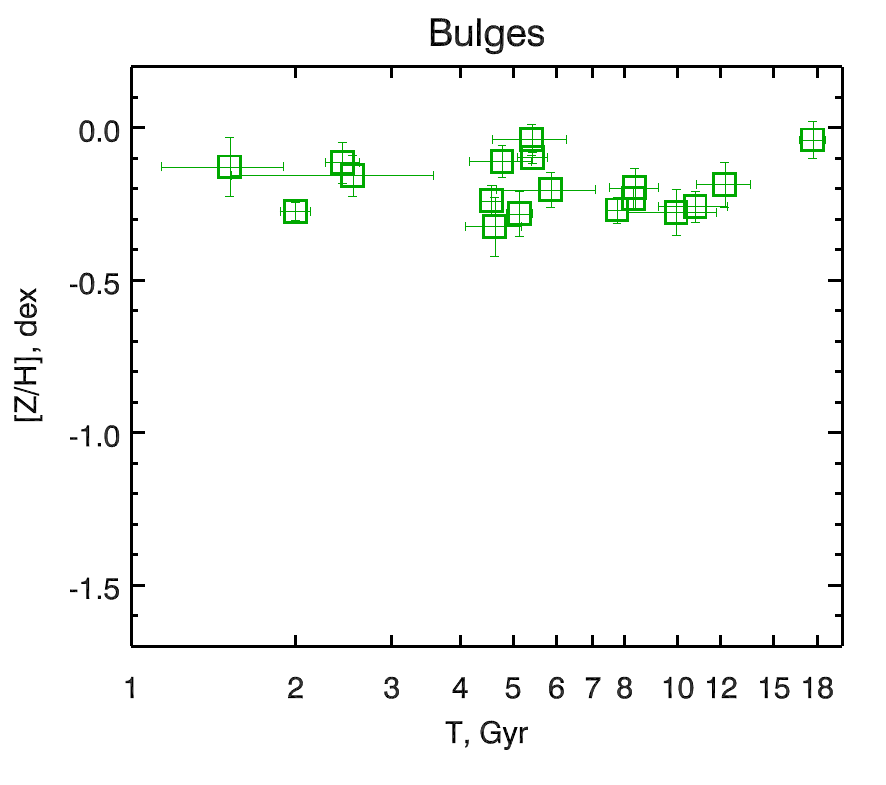}
\includegraphics[width=0.3\textwidth]{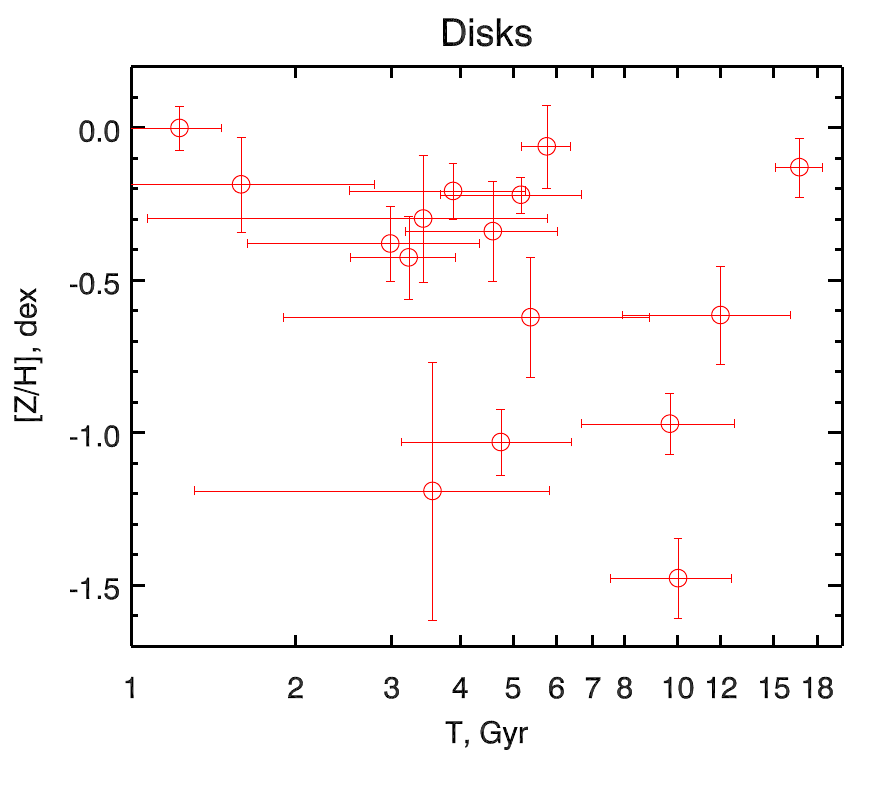}
\includegraphics[width=0.3\textwidth]{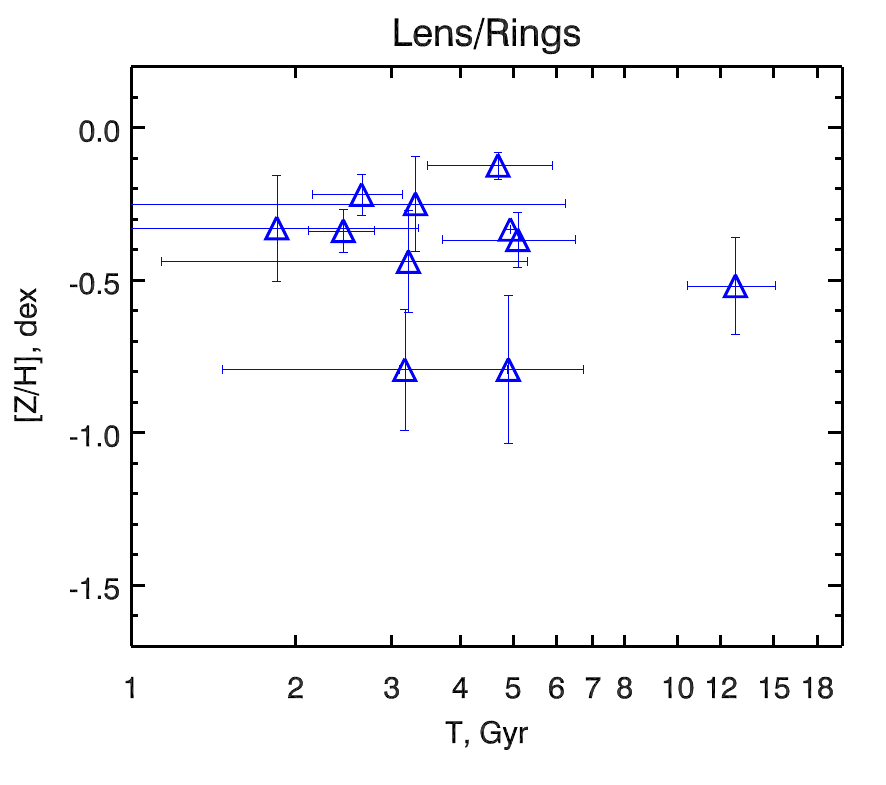}
}
\centerline{
\includegraphics[width=0.3\textwidth]{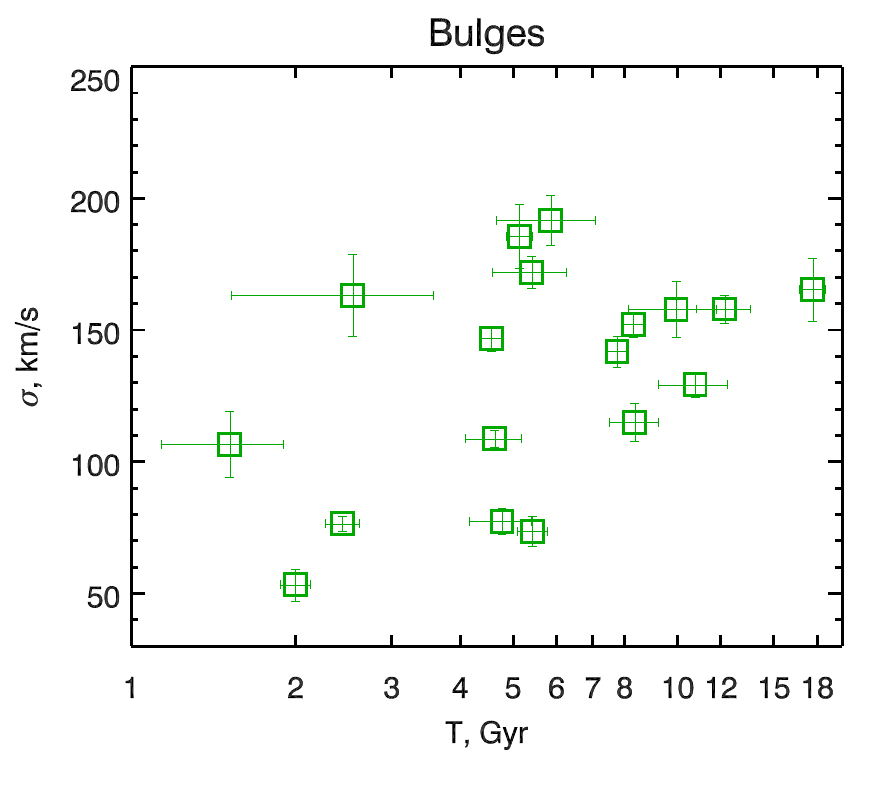}
\includegraphics[width=0.3\textwidth]{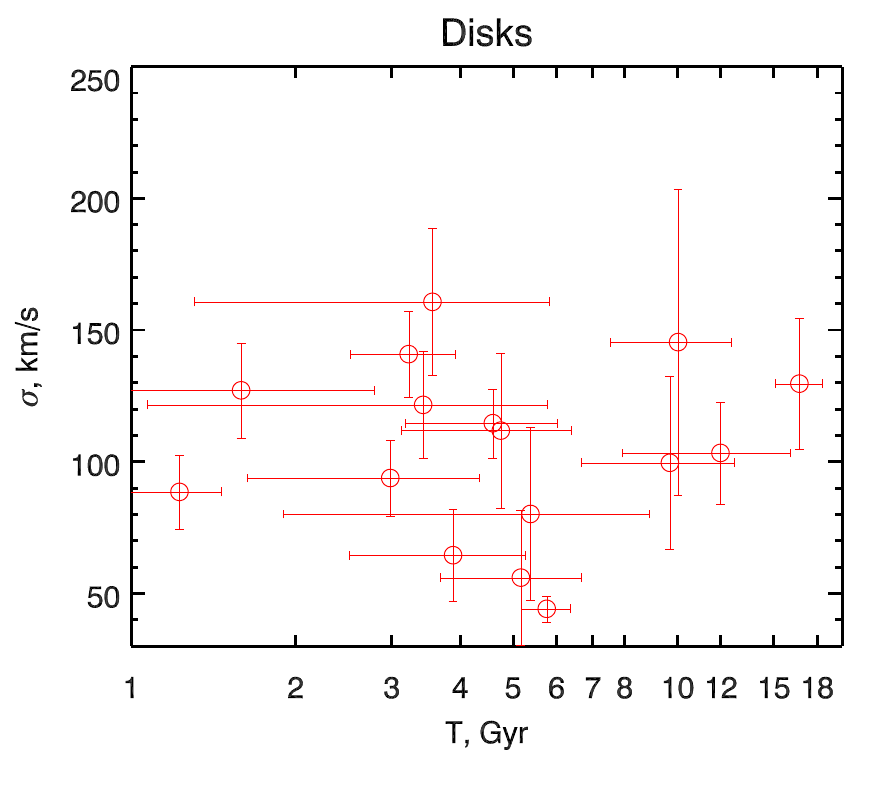}
\includegraphics[width=0.3\textwidth]{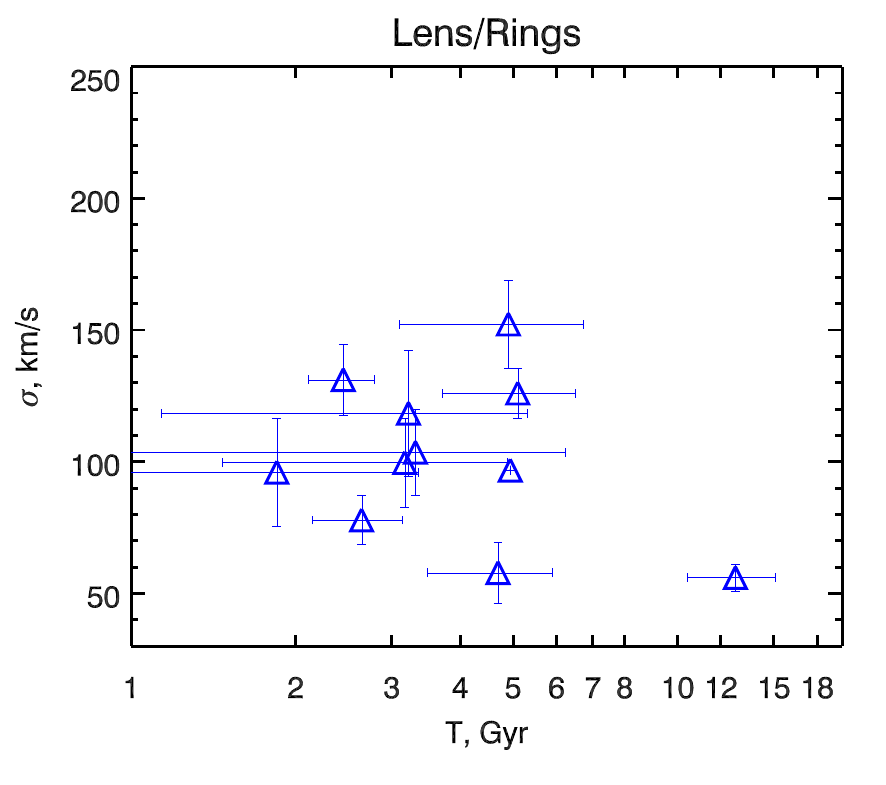}
}
\centerline{
\includegraphics[width=0.3\textwidth]{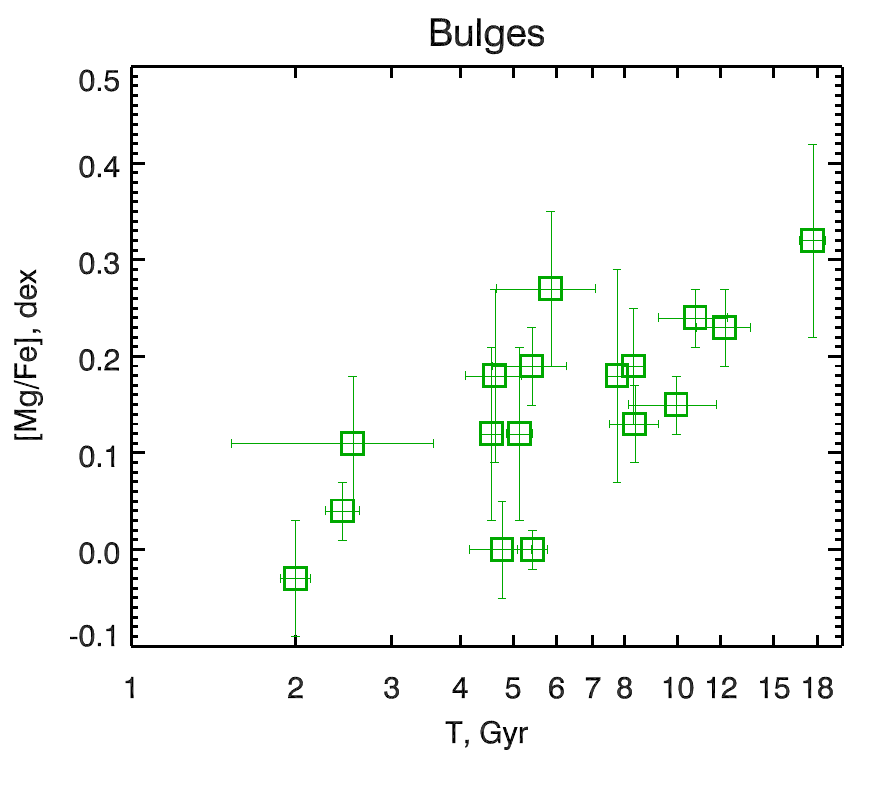}
\includegraphics[width=0.3\textwidth]{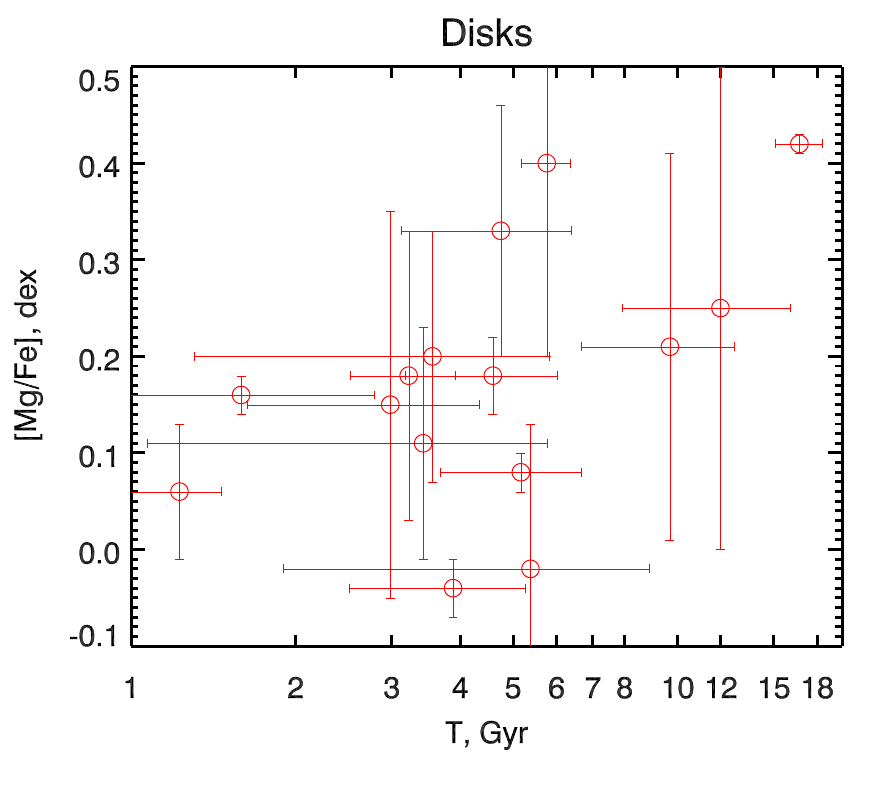}
\includegraphics[width=0.3\textwidth]{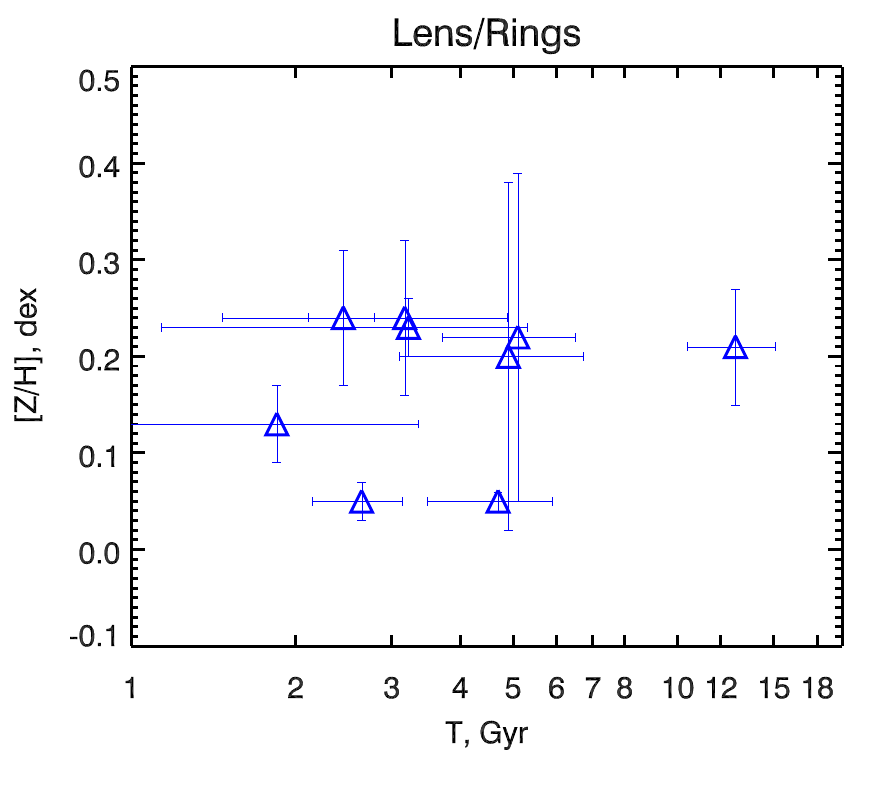}
}
\centerline{
\includegraphics[width=0.3\textwidth]{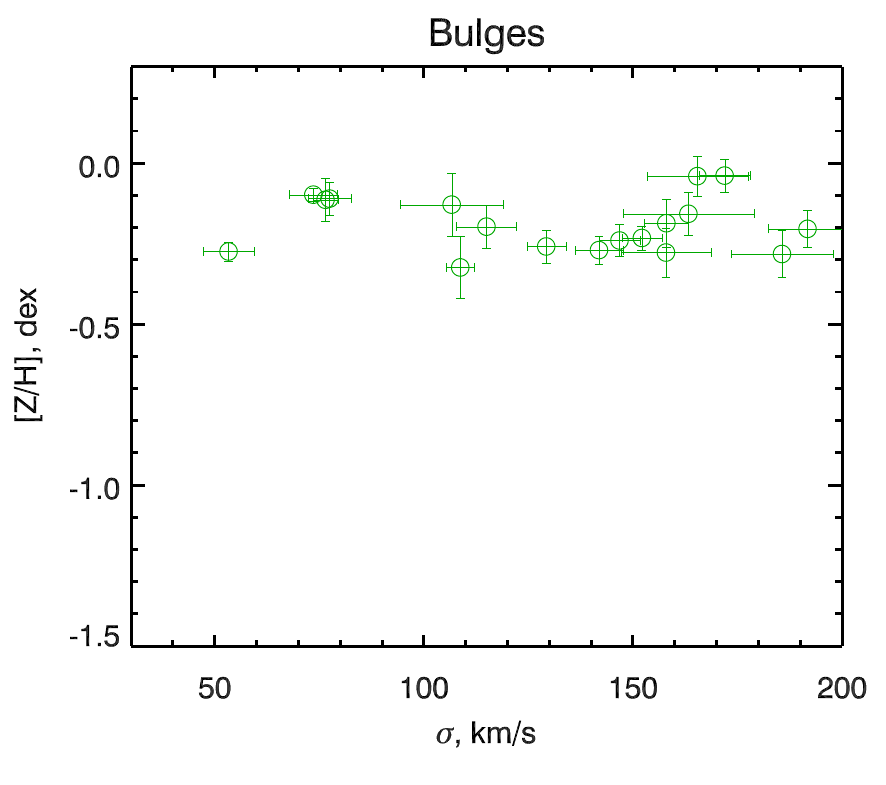}
\includegraphics[width=0.3\textwidth]{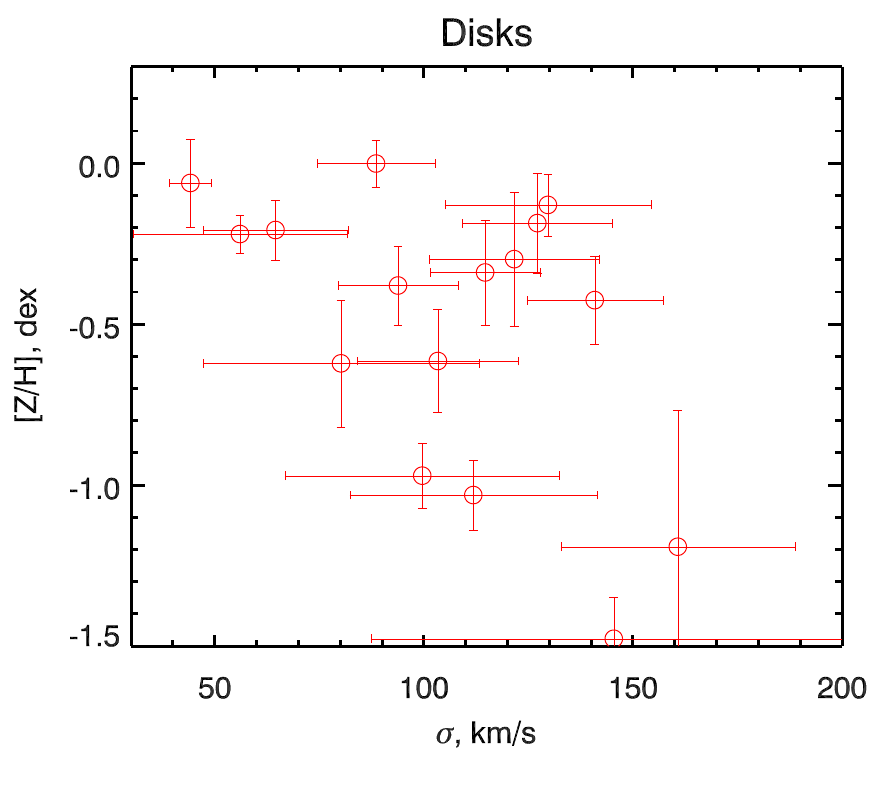}
\includegraphics[width=0.3\textwidth]{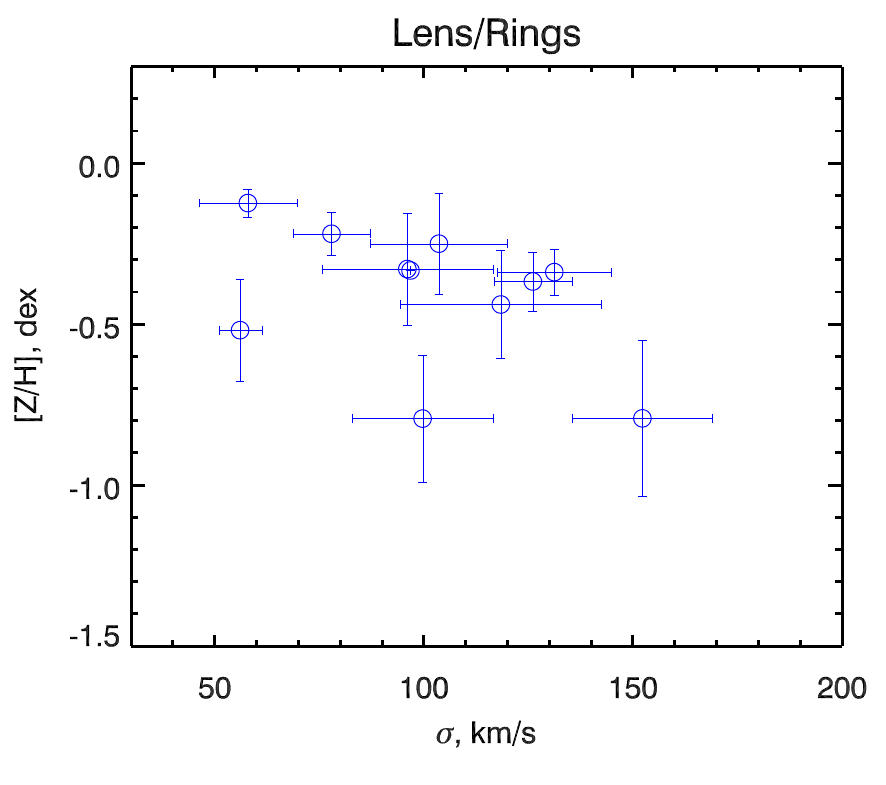}
}
\centerline{
\includegraphics[width=0.3\textwidth]{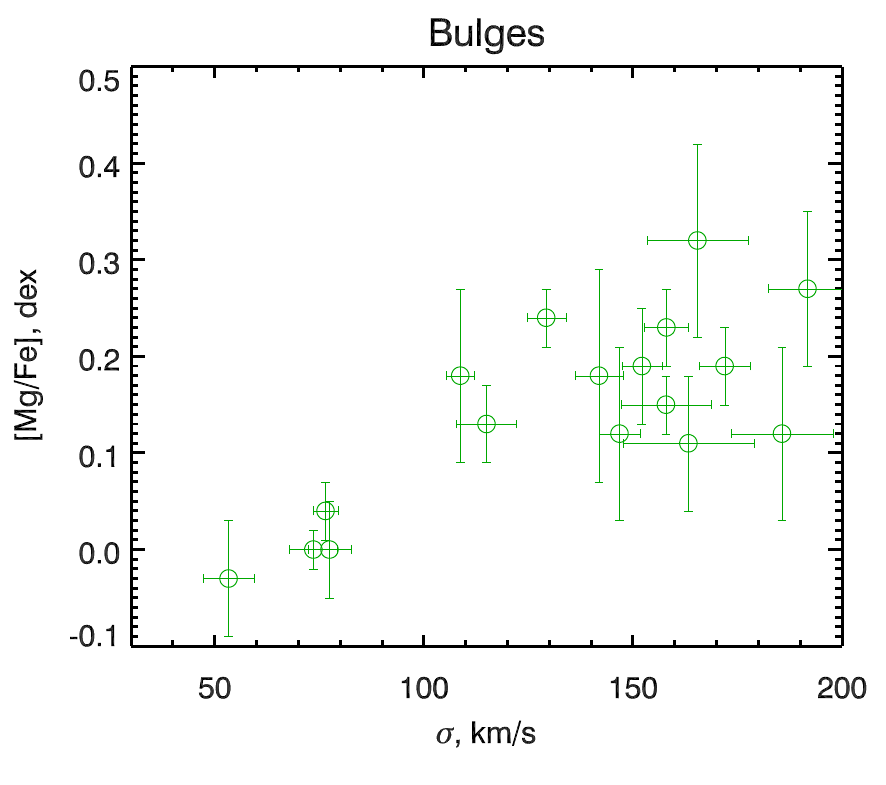}
\includegraphics[width=0.3\textwidth]{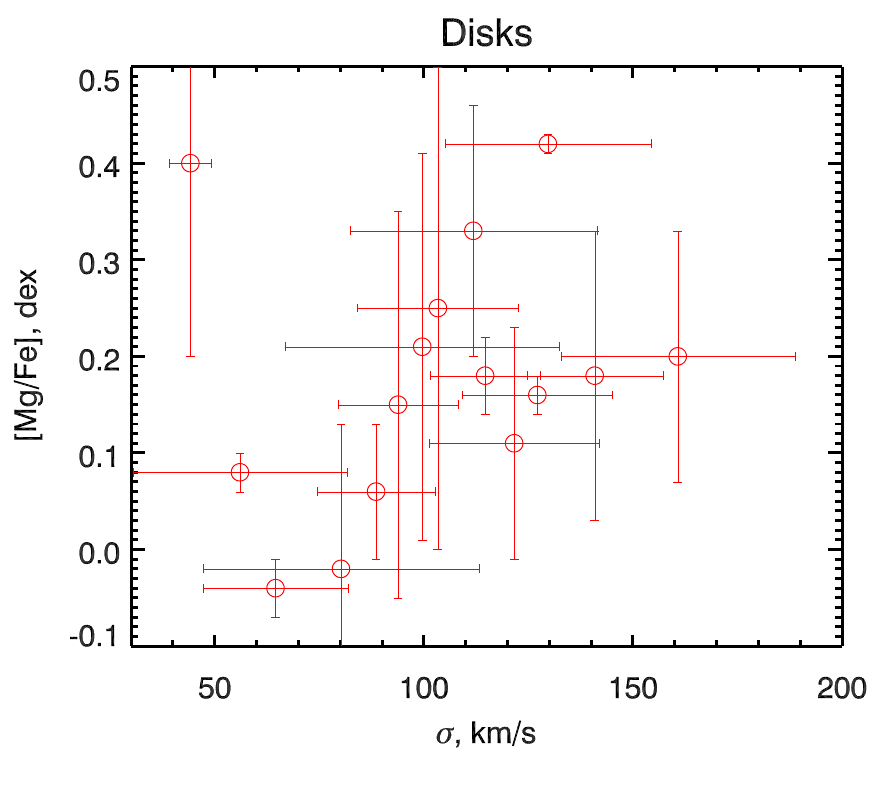}
\includegraphics[width=0.3\textwidth]{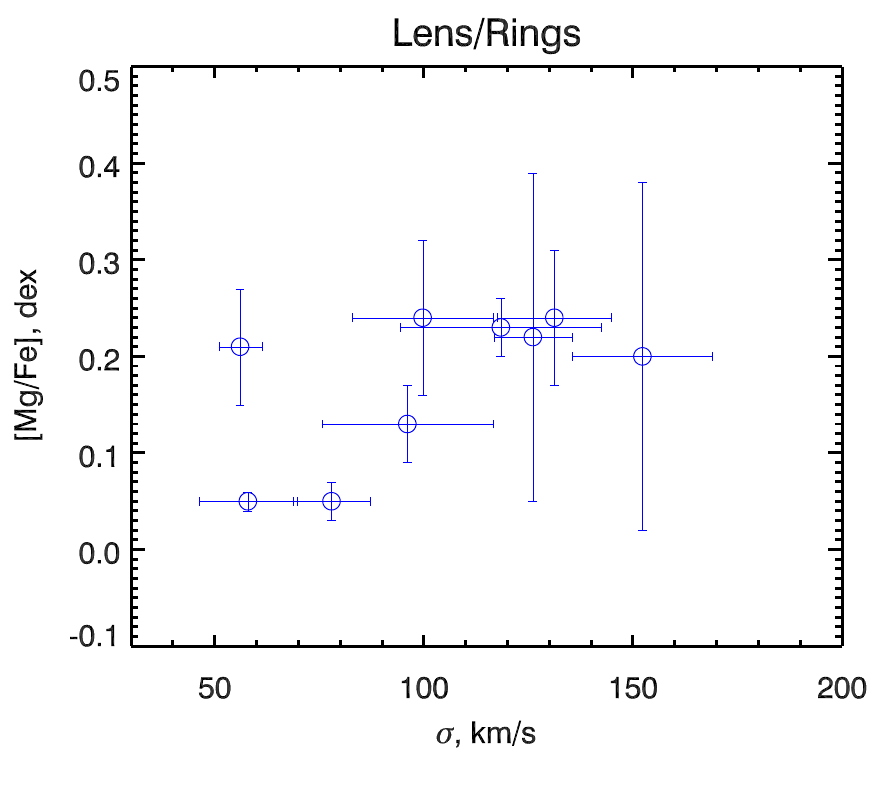}
}
\caption{Parameter dependancies for bulges, disc and lens/ring.}\label{pics_stpop_matrix}
\end{figure*}

\begin{figure*}
\centerline{
\includegraphics[width=0.45\textwidth]{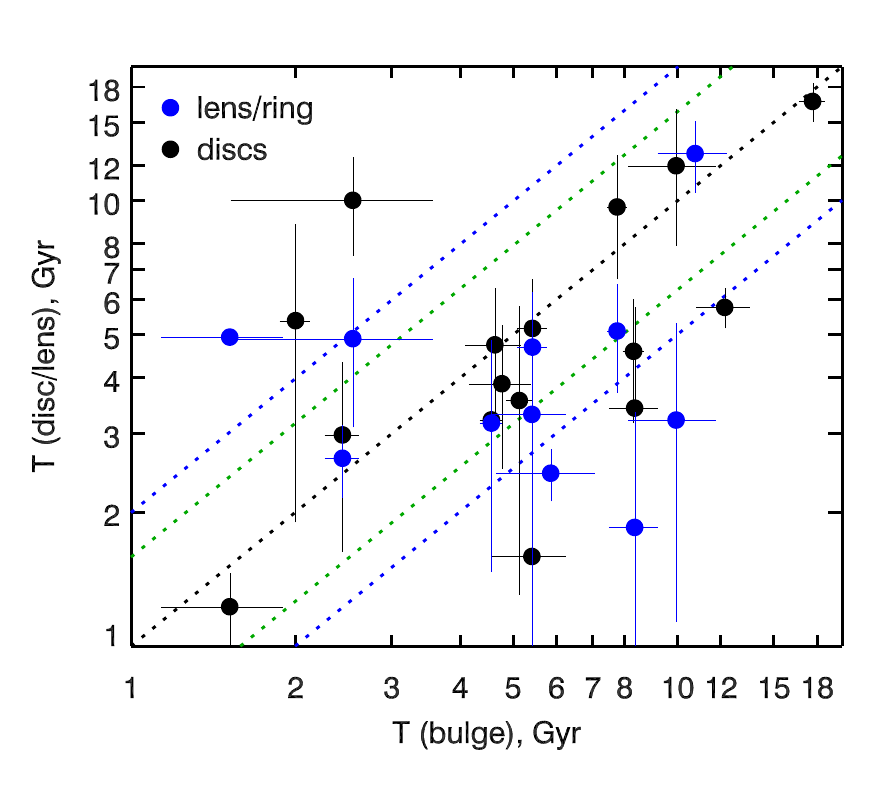}
\includegraphics[width=0.45\textwidth]{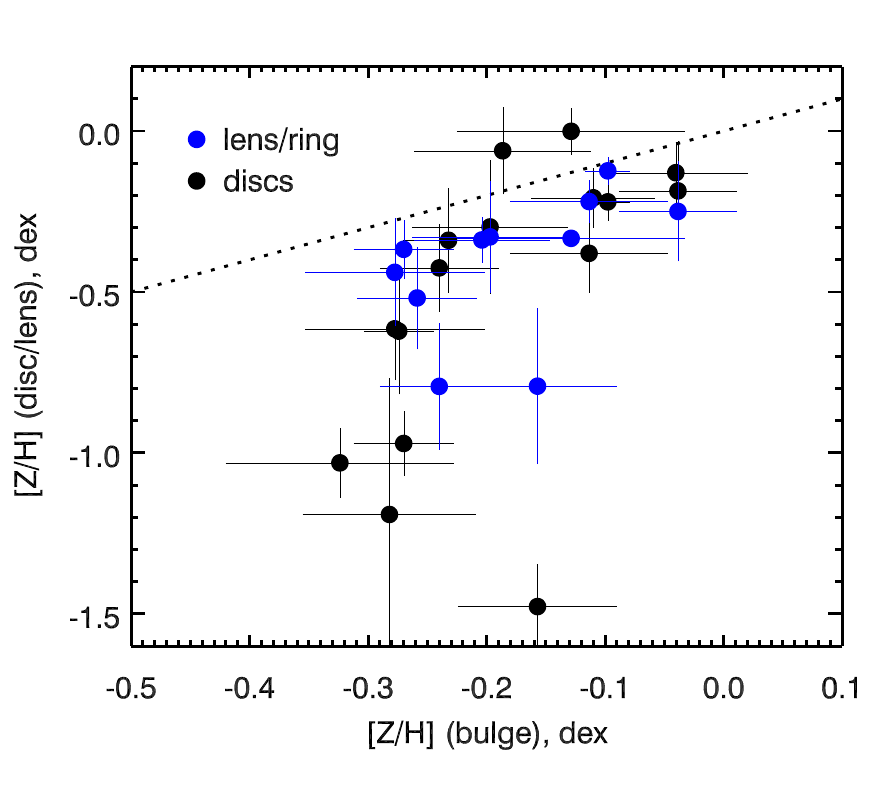}}
\centerline{
\includegraphics[width=0.45\textwidth]{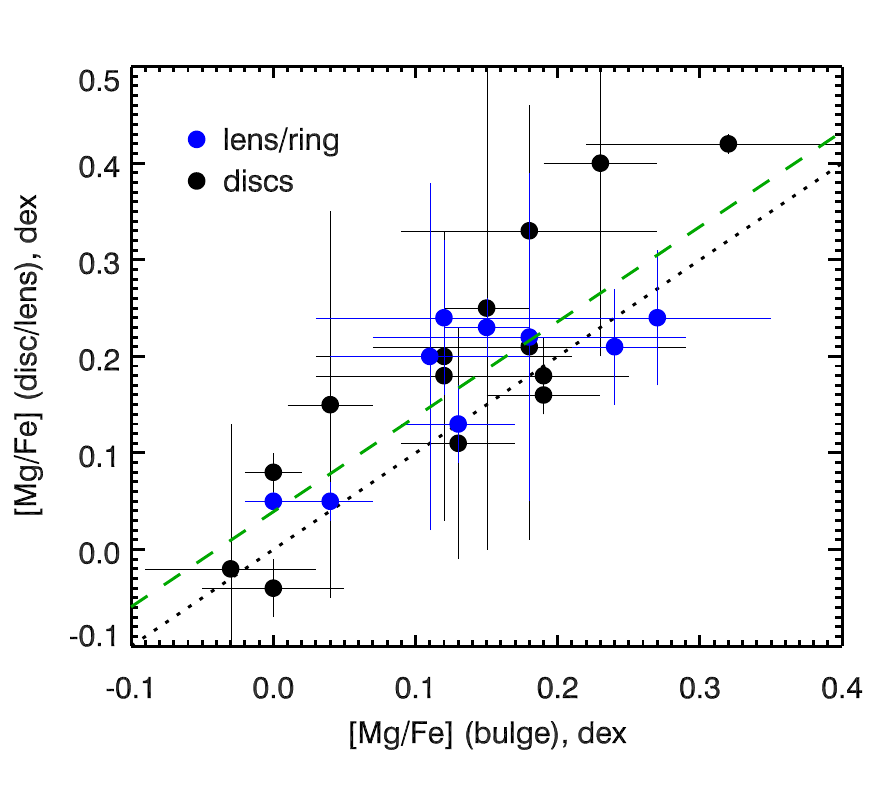}
\includegraphics[width=0.45\textwidth]{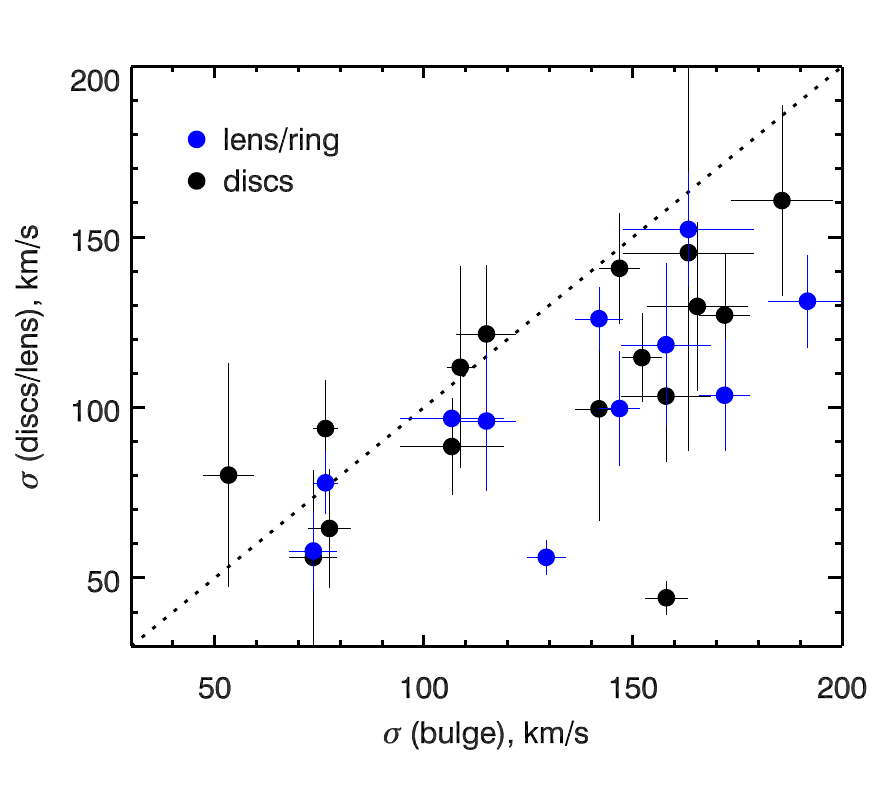}}
\caption{Comparison of stellar population properties between bulges and disc, lens/ring. Dotted line corresponds to equilibrium line. Green dashed line shows a linear fit taking into account error bars along X as well as Y direction.}\label{pics_stpop_bulge_disc}
\end{figure*}

In order to compare stellar population properties of different galactic components (bulges, discs, lens/ring) we have averaged radial profiles in the region where each component have dominant contribution. We computed azimuthally averaged surface brightness profiles using photometric data from SDSS (DR9), 2MASS and direct imaging in 6-m telescope and determined the dominance regions for every structural component of galaxy. We marked out disc regions where surface brightness profiles are exactly fitted by exponential profile.   The averaged stellar parameters are shown in Table~\ref{table_stpop_all}, Figures~\ref{pics_stpop_hist}, \ref{pics_stpop_bulge_disc}, \ref{pics_stpop_matrix} and \ref{pics_stpop_discs_lens} represent stellar population properties and comparison between them for different galactic components.

\begin{table*}
{\small{}
\centerline{
\caption{Averaged stellar population parameters}\label{table_stpop_all}
\begin{tabular}{lccccc}
\hline\hline
Galaxy & N & T, Gyr & [Z/H], dex & [Mg/Fe], dex, & $\sigma$, km/s \\
\hline\hline
\multicolumn{6}{c}{Bulge}\\
\hline
IC 1502  &     11 & $   17.7^{\pm 1.0}$ & $   -0.04^{\pm 0.06}$ & $   0.32^{\pm 0.10}$ & $  165^{\pm 12}$ \\
IC 1608  &      3 & $    4.6^{\pm 0.2}$ & $   -0.24^{\pm 0.05}$ & $   0.12^{\pm 0.09}$ & $  147^{\pm  5}$ \\
IC 3152  &      6 & $    5.1^{\pm 0.3}$ & $   -0.28^{\pm 0.07}$ & $   0.12^{\pm 0.09}$ & $  186^{\pm 12}$ \\
NGC 16   &     10 & $    5.4^{\pm 0.8}$ & $   -0.04^{\pm 0.05}$ & $   0.19^{\pm 0.04}$ & $  172^{\pm  6}$ \\
NGC 1211 &      4 & $    2.5^{\pm 1.0}$ & $   -0.16^{\pm 0.07}$ & $   0.11^{\pm 0.07}$ & $  163^{\pm 16}$ \\
NGC 2350 &      8 & $    1.5^{\pm 0.4}$ & $   -0.13^{\pm 0.10}$ &     --               & $  107^{\pm 12}$ \\
NGC 2917 &      4 & $    5.9^{\pm 1.2}$ & $   -0.20^{\pm 0.06}$ & $   0.27^{\pm 0.08}$ & $  192^{\pm  9}$ \\
NGC 3098 &     10 & $    5.4^{\pm 0.3}$ & $   -0.10^{\pm 0.02}$ & $   0.00^{\pm 0.02}$ & $   74^{\pm  6}$ \\
NGC 3248 &     10 & $    4.8^{\pm 0.6}$ & $   -0.11^{\pm 0.05}$ & $   0.00^{\pm 0.05}$ & $   77^{\pm  5}$ \\
NGC 4240 &      6 & $    4.6^{\pm 0.5}$ & $   -0.32^{\pm 0.10}$ & $   0.18^{\pm 0.09}$ & $  109^{\pm  3}$ \\
NGC 6010 &      3 & $    8.3^{\pm 0.4}$ & $   -0.23^{\pm 0.04}$ & $   0.19^{\pm 0.06}$ & $  152^{\pm  5}$ \\
NGC 6615 &      8 & $   10.8^{\pm 1.6}$ & $   -0.26^{\pm 0.05}$ & $   0.24^{\pm 0.03}$ & $  129^{\pm  5}$ \\
NGC 6654 &      9 & $   12.2^{\pm 1.4}$ & $   -0.19^{\pm 0.07}$ & $   0.23^{\pm 0.04}$ & $  158^{\pm  5}$ \\
NGC 6798 &      8 & $    8.3^{\pm 0.9}$ & $   -0.20^{\pm 0.07}$ & $   0.13^{\pm 0.04}$ & $  115^{\pm  7}$ \\
NGC 7351 &      8 & $    2.0^{\pm 0.1}$ & $   -0.27^{\pm 0.03}$ & $  -0.03^{\pm 0.06}$ & $   53^{\pm  6}$ \\
UGC 4551 &      8 & $    9.9^{\pm 1.8}$ & $   -0.28^{\pm 0.08}$ & $   0.15^{\pm 0.03}$ & $  158^{\pm 10}$ \\
UGC 9519 &      8 & $    2.4^{\pm 0.2}$ & $   -0.11^{\pm 0.07}$ & $   0.04^{\pm 0.03}$ & $   76^{\pm  3}$ \\
UGC 9980 &      4 & $    7.7^{\pm 0.3}$ & $   -0.27^{\pm 0.04}$ & $   0.18^{\pm 0.11}$ & $  142^{\pm 6}$ \\
\hline
\multicolumn{6}{c}{Disc}\\
\hline
IC 1502  &     11 & $   16.7^{\pm 1.7}$ & $   -0.13^{\pm 0.10}$ & $    0.42^{\pm 0.01}$ & $  130^{\pm 25}$ \\
IC 1608  &      8 & $    3.2^{\pm 0.7}$ & $   -0.43^{\pm 0.14}$ & $    0.18^{\pm 0.15}$ & $  141^{\pm 16}$ \\
IC 3152  &      8 & $    3.6^{\pm 2.3}$ & $   -1.19^{\pm 0.42}$ & $    0.20^{\pm 0.13}$ & $  161^{\pm 28}$ \\
NGC 16   &     18 & $    1.6^{\pm 1.2}$ & $   -0.19^{\pm 0.16}$ & $    0.16^{\pm 0.02}$ & $  127^{\pm 18}$ \\
NGC 1211 &      2 & $   10.0^{\pm 2.5}$ & $   -1.48^{\pm 0.13}$ &      --               & $  145^{\pm 58}$ \\
NGC 2350 &     14 & $    1.2^{\pm 0.2}$ & $   -0.00^{\pm 0.07}$ & $    0.06^{\pm 0.07}$ & $   89^{\pm 14}$ \\
NGC 2917 &      0 &      -- &      -- &      -- & -- \\
NGC 3098 &     19 & $    5.2^{\pm 1.5}$ & $   -0.22^{\pm 0.06}$ & $    0.08^{\pm 0.02}$ & $   56^{\pm 26}$ \\
NGC 3248 &     31 & $    3.9^{\pm 1.4}$ & $   -0.21^{\pm 0.09}$ & $   -0.04^{\pm 0.03}$ & $   65^{\pm 17}$ \\
NGC 4240 &      5 & $    4.8^{\pm 1.6}$ & $   -1.03^{\pm 0.11}$ & $    0.33^{\pm 0.13}$ & $  112^{\pm 30}$ \\
NGC 6010 &     11 & $    4.6^{\pm 1.4}$ & $   -0.34^{\pm 0.16}$ & $    0.18^{\pm 0.04}$ & $  115^{\pm 13}$ \\
NGC 6615 &      0 &      -- &      -- &      -- & -- \\
NGC 6654 &      3 & $    5.8^{\pm 0.6}$ & $   -0.06^{\pm 0.14}$ & $    0.40^{\pm 0.20}$ & $   44^{\pm  5}$ \\
NGC 6798 &     22 & $    3.4^{\pm 2.4}$ & $   -0.30^{\pm 0.21}$ & $    0.11^{\pm 0.12}$ & $  122^{\pm 20}$ \\
NGC 7351 &      7 & $    5.4^{\pm 3.5}$ & $   -0.62^{\pm 0.20}$ & $   -0.02^{\pm 0.15}$ & $   80^{\pm 33}$ \\
UGC 4551 &      1 & $   12.0^{\pm 0.0}$ & $   -0.61^{\pm 0.00}$ & $    0.25^{\pm 0.25}$ & $  103^{\pm  0}$ \\
UGC 9519 &      4 & $    3.0^{\pm 1.4}$ & $   -0.38^{\pm 0.12}$ & $    0.15^{\pm 0.20}$ & $   94^{\pm 14}$ \\
UGC 9980 &      6 & $    9.7^{\pm 3.0}$ & $   -0.97^{\pm 0.10}$ & $    0.21^{\pm 0.20}$ & $  100^{\pm 33}$ \\
\hline
\multicolumn{6}{c}{Lens/Ring}\\
\hline
IC 1502  &      0 &      -- &      -- &      -- & -- \\
IC 1608  &     10 & $    3.2^{\pm 1.7}$ & $   -0.79^{\pm 0.20}$ & $    0.24^{\pm 0.08}$ & $  100^{\pm 17}$ \\
IC 3152  &      0 &      -- &      -- &      -- & -- \\
NGC 16 &     16 & $      3.3^{\pm 2.9}$ & $   -0.25^{\pm 0.16}$ &      --               & $  104^{\pm 16}$ \\
NGC 1211 &     12 & $    4.9^{\pm 1.8}$ & $   -0.79^{\pm 0.24}$ & $    0.20^{\pm 0.18}$ & $  152^{\pm 17}$ \\
NGC 2350 &      1 & $    4.9^{\pm 0.0}$ & $   -0.33^{\pm 0.00}$ &      --               & $   97^{\pm  0}$ \\
NGC 2917 &     10 & $    2.5^{\pm 0.3}$ & $   -0.34^{\pm 0.07}$ & $    0.24^{\pm 0.07}$ & $  131^{\pm 14}$ \\
NGC 3098 &     13 & $    4.7^{\pm 1.2}$ & $   -0.12^{\pm 0.04}$ & $    0.05^{\pm 0.01}$ & $   58^{\pm 12}$ \\
NGC 3248 &      0 &      -- &      -- &      -- & -- \\
NGC 4240 &      0 &      -- &      -- &      -- & -- \\
NGC 6010 &      0 &      -- &      -- &      -- & -- \\
NGC 6615 &      3 & $   12.8^{\pm 2.4}$ & $   -0.52^{\pm 0.16}$ & $    0.21^{\pm 0.06}$ & $   56^{\pm  5}$ \\
NGC 6654 &      0 &      -- &      -- &      -- & -- \\
NGC 6798 &     20 & $    1.9^{\pm 1.5}$ & $   -0.33^{\pm 0.17}$ & $    0.13^{\pm 0.04}$ & $   96^{\pm 20}$ \\
NGC 7351 &      0 &      -- &      -- &      -- & -- \\
UGC 4551 &     11 & $    3.2^{\pm 2.1}$ & $   -0.44^{\pm 0.17}$ & $    0.23^{\pm 0.03}$ & $  118^{\pm 24}$ \\
UGC 9519 &     23 & $    2.6^{\pm 0.5}$ & $   -0.22^{\pm 0.07}$ & $    0.05^{\pm 0.02}$ & $   78^{\pm  9}$ \\
UGC 9980 &      6 & $    5.1^{\pm 1.4}$ & $   -0.37^{\pm 0.09}$ &  --  & $126^{\pm 9}$ \\
\hline\hline
\end{tabular}
}}
\end{table*}

\subsection{Comparison of stellar populations for bulge, disc, lens/ring.}

The next step is a comparison of stellar population properties of various components of galaxies. The ratio of average ages produces a sequence of major evolutionary stages in the galactic evolution path, and the $\alpha$-element abundance ratio [Mg/Fe] limits the epoch of star formation.

\subsubsection{Bulges vs. discs}
In the standard $\Lambda$CDM cosmology \citep{Blumenthal_1984} galaxies assemble through hierarchical merging and then evolve under a combination of internal and environmental processes. In this paradigm the classical bulges are formed firstly by major merging and then the large-scale disc components are assembled by smoothed cold gas accretion \citep{Governato_2004,Governato_2007}. However, there are observational evidences

However, there are observational evidences contradicting these predictions: the studies of the surface photometry of galaxies consistently demonstrate the correlation between the characteristic scales of bulges and discs, and the presence of this correlation does not depend on the type of bulges -- classical bulge or pseudo-bulges \citep{macarthur03,mendez_abreu,laurikainen_2010}. Thus, the formation of bulge and disc structure of galaxy seems to be synchronized in galaxies.

It is supported by Fig.~\ref{pics_stpop_bulge_disc} (\textit{upper left}). At first sight the dependence of age of discs against bulges looks rather ``smeared'', however the correlation coefficient (Pearson) is equal 0.66. At Fig.~\ref{pics_stpop_bulge_disc} (\textit{upper left}) the green dotted lines denote the area where measurements deviate from bisector less then 0.2 dex (in 1.5 times), blue dotted lines -- less then 0.3 dex (in 3 times). The $\pm0.2$ dex region consist of 63\% (10/16) measurements, the $\pm0.2$ dex region -- 75\% (12/16). So, the comparison of stellar ages of bulges and discs shows a tendency to equal ages of bulge and disc component.

The ages measurements of the isolated lenticular galaxies essentially differ from measurements for galaxies in the dense environment: the points are grouped in the left top corner, over a bisector -- the ages of discs are equal, or older in comparison with the bulge ages \citep{sil_s0}. That difference between group/cluster galaxies and galaxies in sparse environment is expected. Because all physical mechanisms of external impact related to dense environment, both gravitational and gas-dynamic, lead to gas flows down to the galaxy center and provoke the subsequent star formation bursts in the center and  bulge rejuvenation \citep{bekki_couch,kronberger}. 

It is interesting that the alpha abundance ratio is identical for bulges and discs (see. Fig.~\ref{pics_stpop_bulge_disc}, \textit{right}). We calculated a linear fit ($[Mg/Fe]_{disc}=a+b[Mg/Fe]_{bulge}$) taking into account error bars along both axes and found  no significant deviations of zero-point from zero level as well as slope of dependence:  $a=0.04\pm0.05$, $b=0.9\pm0.3$. The strong correlation ($R_P=0.88$) implies that the star formation fades quickly or continues during billions years simultaneously both in bulges and discs. This result allows us to strengthen the thesis about synchronous formation of bulges and discs: the star formation starts and stops quasi-simultaneously in both components. The stellar metallicity of discs is lower than bugles. That possibly means that the low-metallicity gas accretes generally onto the disc where gas is enriched by metals and goes down to the bulge region.

\begin{figure}
\includegraphics[width=\textwidth]{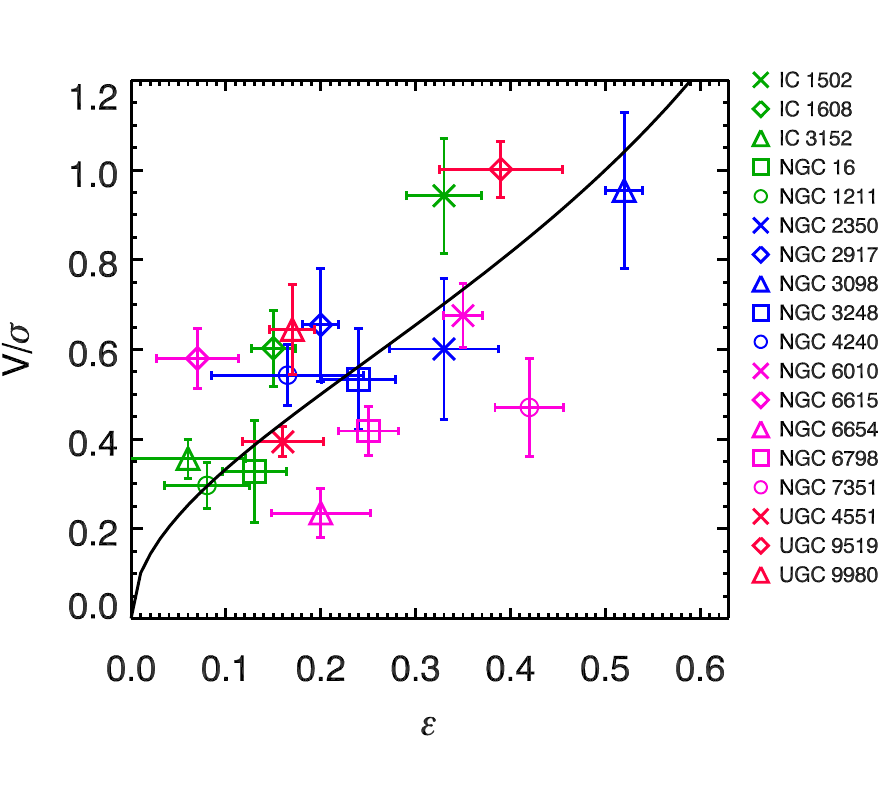}
\caption{Kormendy diagram: ratio between velocity rotation and stellar velocity dispersion $V/\sigma$ against isophote ellipticity $\epsilon$. Solid line correspond to isotropic rotationally supported spheroids.}\label{pics_stpop_Kormendy_bulges}
\end{figure}

Based on Kormendy diagram (dependence of $V/\sigma$ ratio on isophotal ellipticity, Fig.~\ref{pics_stpop_Kormendy_bulges}) and relationship between bulge and disc velocity dispersion it appeared equally pseudo-bulges and dynamically heated classical bulges. It confirms once again that bulges in the lenticular galaxies are very various on luminosity and a contribution to the total galaxy mass as well as by origin and evolution. It is true also if we consider the \textit{isolated} S0 galaxies, for which the impact of an environment on evolution seems to minimized.

\begin{figure*}
\centerline{
\includegraphics[width=0.45\textwidth]{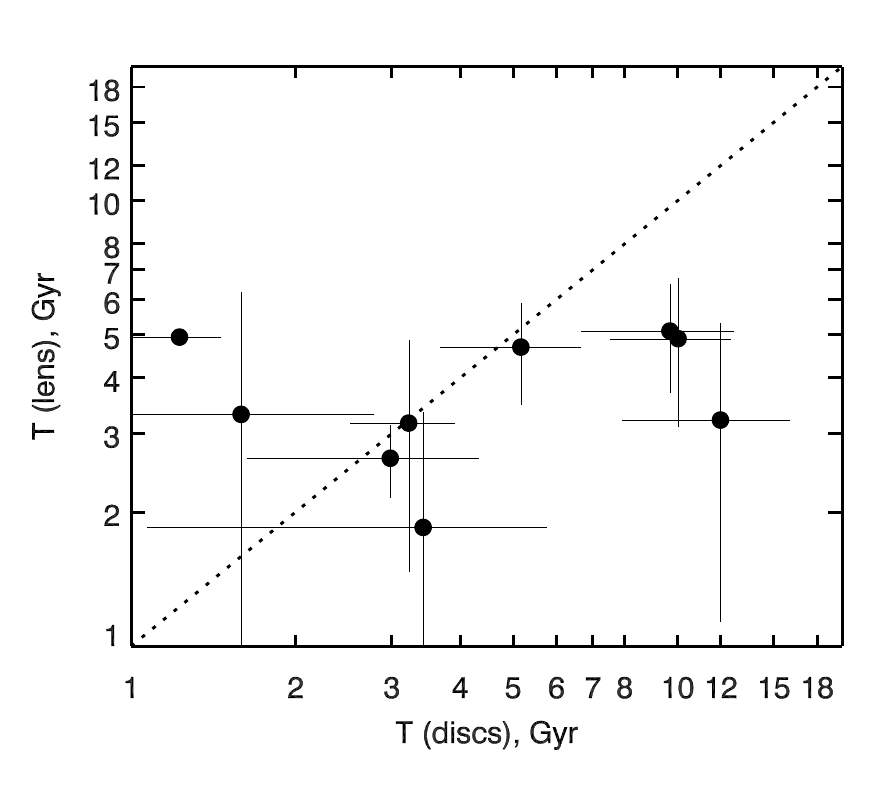}
\includegraphics[width=0.45\textwidth]{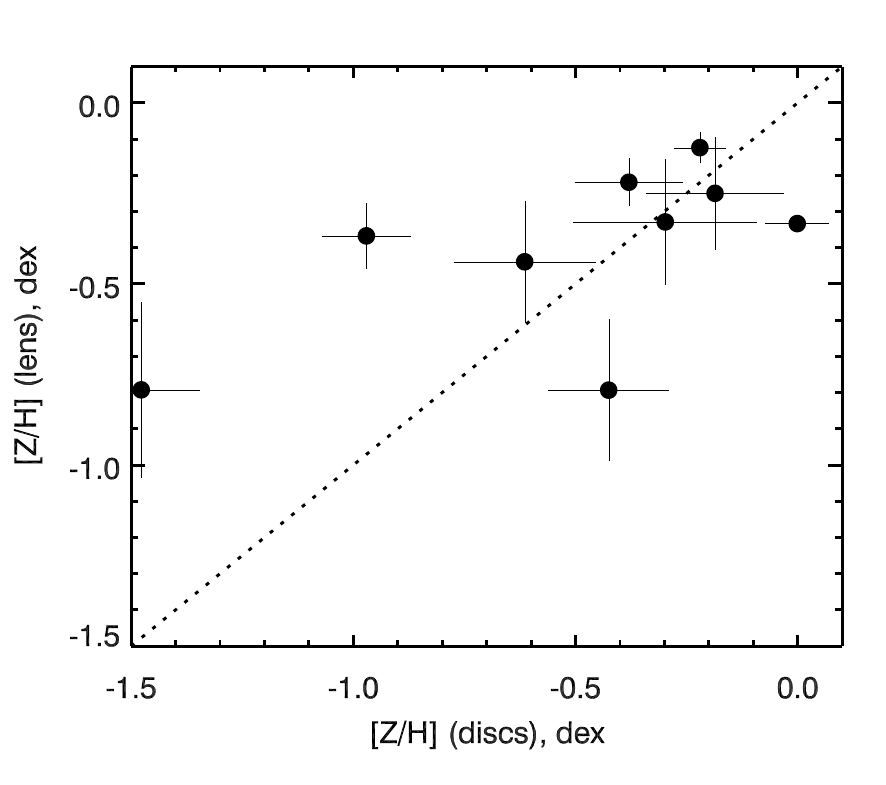}}
\centerline{
\includegraphics[width=0.45\textwidth]{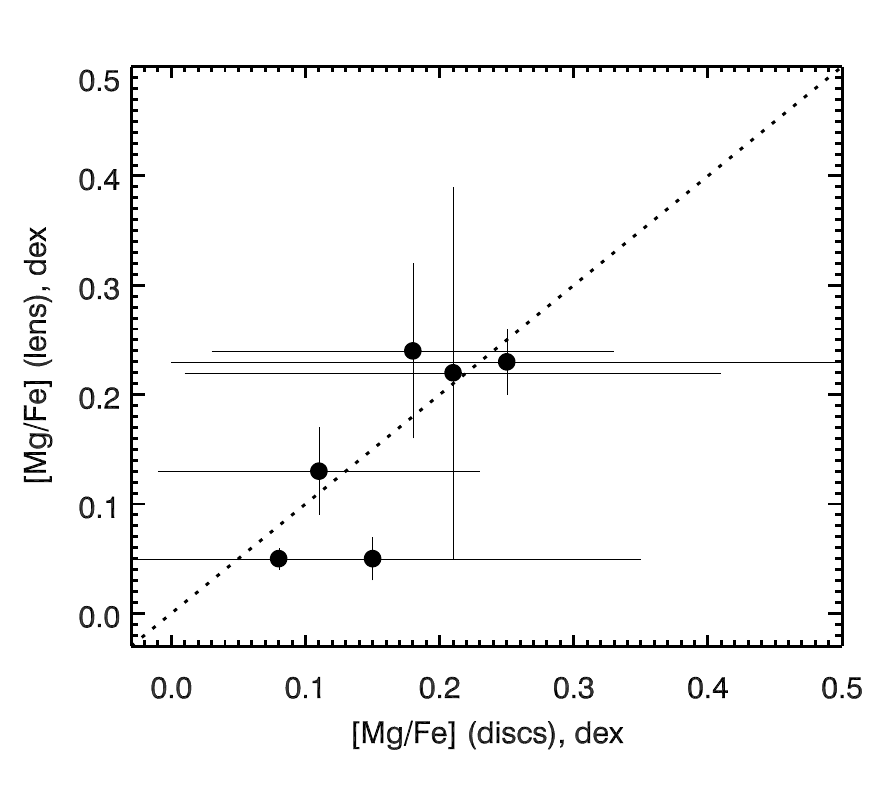}
\includegraphics[width=0.45\textwidth]{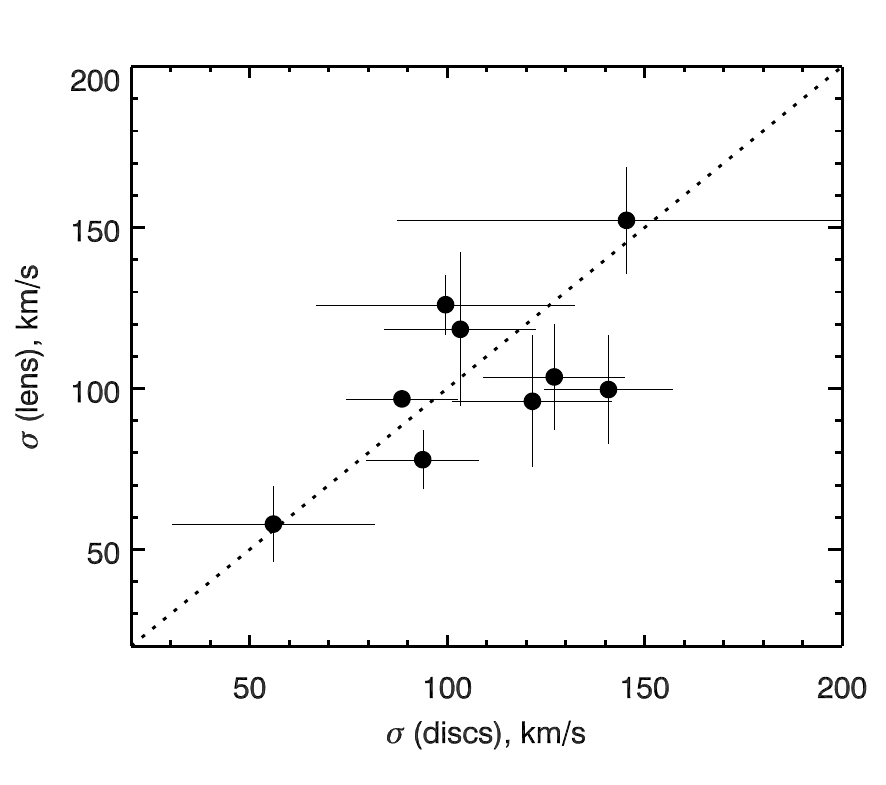}
}
\caption{Comparison of stellar population properties between discs and lenses/rings.}\label{pics_stpop_discs_lens}
\end{figure*}

\subsubsection{Disc substructures: lenses and rings}
The lenses in S0 galaxies are considered genetically connected with starforming rings in past. Still the prevailed point of view that stellar population in lenses are old, and dynamically they are hot \citep{kormendy_n1553,lauri13}. However, this opinion is based on studying of a single objects. We were able to determine stellar population properties in 10 lenses/rings in our galaxies. On average, the lenses/rings show the same velocity dispersion as the discs (Fig.~\ref{pics_stpop_discs_lens} \textit{right bottom}). Except NGC~6615 ($T_{lens}=13$ Gyr), all measurements in the lenses/rings are not so old, but have intermediate ages in the narrow range $2-5$ Gyr ($z=0.3-0.5$), thus discs ages cover significantly wider range. It means that the rings formation epoch isn't connected with formation of discs. On average, the chemical abundances ([Z/H], [Mg/Fe]) in the lenses and rings are the same as in discs, except only NGC~1211 where disc is extremely metal-poor.

\begin{figure*}
\centerline{
\includegraphics[width=0.45\textwidth]{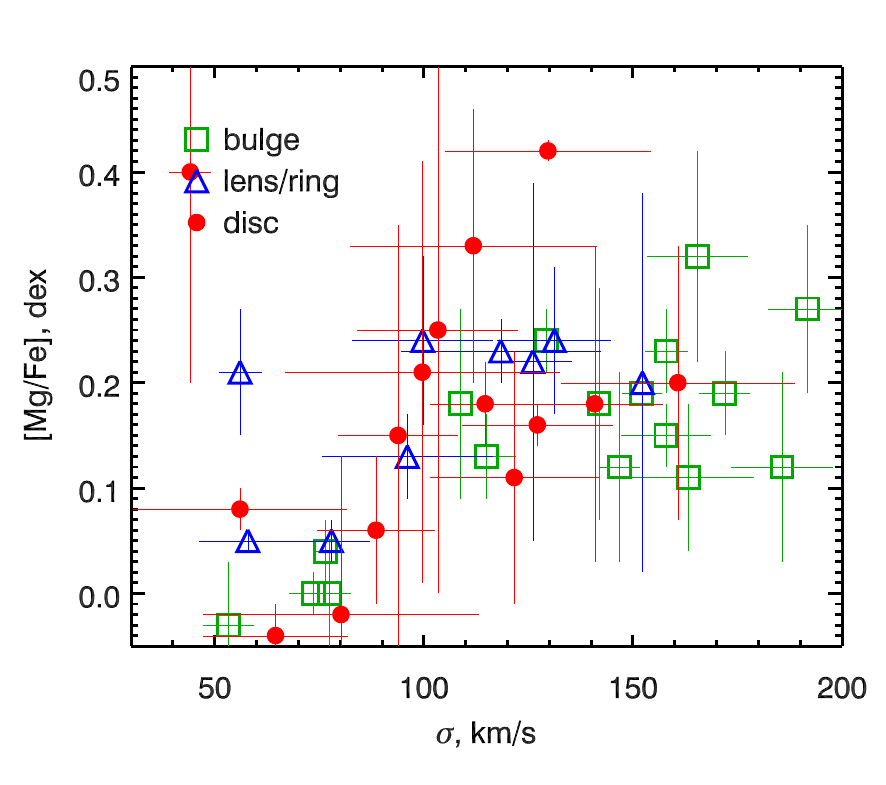}
\includegraphics[width=0.45\textwidth]{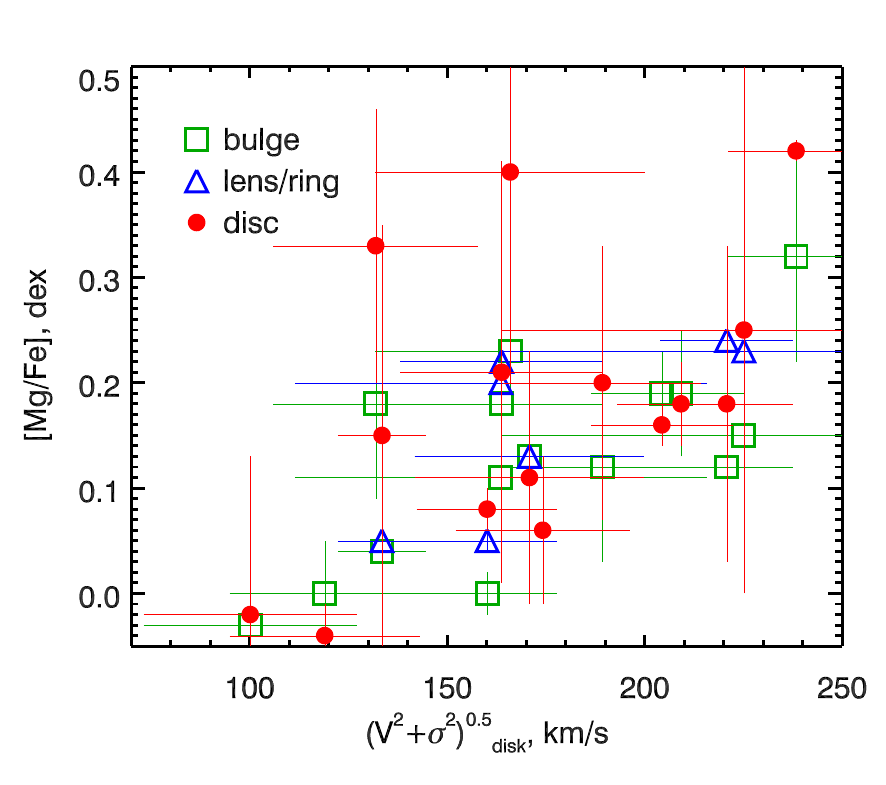}}
\centerline{
\includegraphics[width=0.45\textwidth]{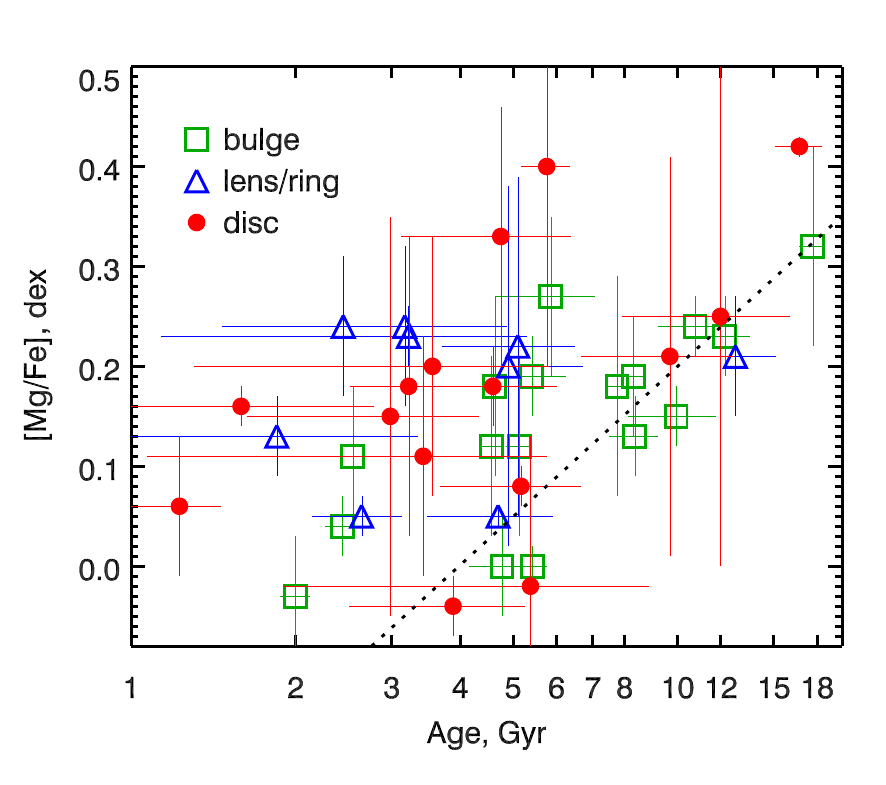}
\includegraphics[width=0.45\textwidth]{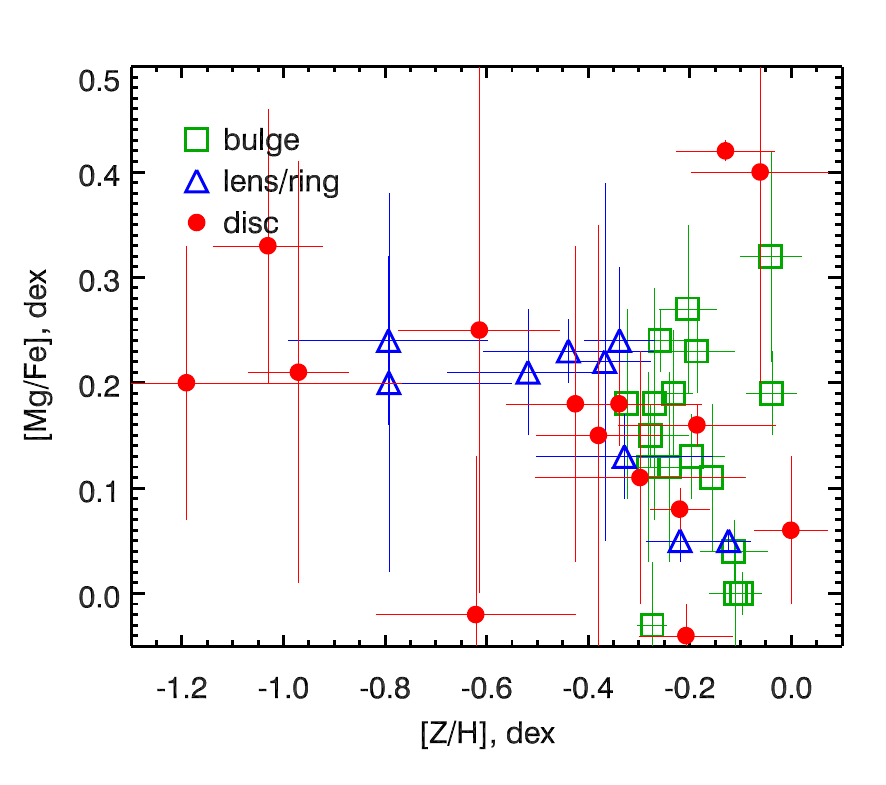}}
\caption{The alpha-abundance ([Mg/Fe]) against velocity dispersion, dynamical parameter $(V^2+\sigma^2)^{0.5}$, ages and metallicity for bulges, discs, lenses/rings.}
\label{plot_stpop_mgfe}
\end{figure*}

\subsection{Epoch and duration of formation stellar content}
The comparison between the alpha element abundance ratio \mgfe, characterized duration of last star formation burst, and other stellar population properties for galaxy components are shown at Fig.~\ref{plot_stpop_mgfe}. And again we see synchronism of bulge evolution and discs. The correlation of the relation of [Mg/Fe] with velocity dispersion of the stars, characterizing the local density of gravitating mass, is known long ago for elliptical galaxies and for bulges \cite{trager}; it is considered as the evidence for relationship of star formation efficiency and depth of potential well. However, Fig.~\ref{plot_stpop_mgfe} (\textit{left}) and Fig.~\ref{pics_stpop_matrix} (\textit{bottom row}), for the first time we see that this correlation is not only for bulges, but also for discs. By analogy it is possible to assume that the accretion rates of external gas are higher when the local potential well in the plane of disc is deeper, and higher rate of accretion provides more effective star formation.

At Fig.~\ref{plot_stpop_mgfe} (\textit{center}) we compare element ratios Mg/Fe with stellar ages. We plot dotted envelop line; the points concentrate near this line or on the left of this line. Probably it is due to the fact that galaxies (and subcomponents) which have begun the formation at the same time in the early Universe, on $z=2-3$, have stopped it in different way. Galaxies, which finished star formation quickly, reveal average older age and higher [Mg/Fe] abundance ratio ; galaxies, where star formation continues during many Gyr's, decrease their [Mg/Fe] abundance ratio down to solar value.
At Fig.~\ref{plot_stpop_mgfe} (\textit{center}) there are cloud of points on left of dotted line consisted of bulge and disc measurements. These are stellar systems in which the last event of a star formation took place \textit{later}, than at the main population of galaxies. Indeed, in order to have ages of $1.5-3$ Gyr and [Mg/Fe]=+0.2, the galaxy needs to experience $1.0-1.5$ billion years of an active star formation significantly later then $z=0.5$. Thus, it turns out that star formation events in the isolated lenticular galaxies can happen at different times and have different duration.

\subsection{Formation scenario}
The results presented in this section confirmed the strong influence environment on evolution and stellar population properties of galaxies: there is no allocated epoch of formation of structural components, they can grow their stellar population on $z>2$ as well as only billion years ago. Our findings support the idea that the dispersion of average ages in discs of S0 galaxies increases with decrease with environment density of galaxies \citep{sil_s0} and reaches a maximum among the isolated galaxies.

On what the morphological type of disc galaxy in sparse environment depends? Why it can be lenticular of spiral galaxy at current epoch? Everything agrees on a mode of external cold gas accretion  which usually feeds a star formation
in discs of spiral galaxies throughout many billions years. The gas accretion  happens quite possibly by stochastic mode.

Recent work by \citet{kara11} on search of dwarf satellites near isolated galaxies reveals curious statistical feature. The radial velocity difference between satellites and host galaxy is higher for lenticular galaxies then for spirals, moreover lenticular galaxies have no satellites with velocity difference $<50$ km s$^{-1}$. Probably this dynamic feature indicates that the satellites of lenticulars galaxies can't fall onto the host galaxy, whereas it is quite enough dynamic friction in order to provide satellite accretion onto spirals. In other words, orbital configurations of dwarf satellites of a galaxy are formed in a stochastic way. Disk galaxies possessing dynamically cold system of satellites will provide themselves by gas ``fuel'' for star formation and subsequent transformation into spiral galaxy. The main physical processes for transformation are minor merging and tidal accretion. If the galaxy can't provide itself by ``fuel'' for star formation due to dynamical hot state of satellite system or due to absence of dwarf which can be fall onto the galaxy, then the galaxy will become lenticular. 


\section{Ionized gas}
We found large-scale ionized gas structures in most objects of our sample of galaxies. For such objects we obtained pure emission-line spectrum by subtracting the stellar contribution (i.e., the best-fitting stellar population model) from the observed spectrum. This step provided a pure emission spectrum uncontaminated by absorption lines of the stellar component that is especially important for the Balmer lines. Then we fitted emission lines with Gaussians pre-convolved with the instrumental LSF in order to determine the LOS velocities of the ionized gas and emission-line fluxes.  

The line fluxes have been corrected for the internal interstellar extinction as well as for the Galactic extinction according to \citet{Schlegel_1998}. The color excess $E(B-V)$ corresponding to the internal dust reddening was determined from the Balmer decrement using the theoretical line ratios F(\Ha)/F(\Hb) = 2.87 for the electron temperature Te = 10,000 K \citep{Osterbrock_2006} and the parameterized extinction curve \citep{Fitzpatrick_1999}. We plotted our measurements of emission lines onto the classical excitation-type BPT (Baldwin, Phillips \& Terlevich) diagnostic diagrams \cite{BPTdiag} to identify the gas ionization source. 

The results and discussion for the SCORPIO part of our sample are presented in \citet{ilg_gas,katkov_astro-ph_ilg_gas}. Here we repeat conclusions based on full SCORPIO+SALT sample of galaxies.

\subsection{Statistics of ionized gas decouplings}

Based on analysis of long-slit spectroscopic data obtained at 6-m BTA telescope as well as at SALT telescope, we found that 13 out of 18 ($72\pm11\%$) observed lenticular galaxies reveal extended emission lines; and among those, 6 galaxies ($46\pm14$ \%) demonstrate decoupled gas kinematics respect to their stellar  discs.

By comparing our results on the frequency of extended ionized gas discs in S0 galaxies with the earlier statistics, we see full agreement: \citet{kuijken_fisher} found ionized gas in 17 of 29 S0s studied, so their fraction of gas-rich S0s is about $58\pm9\%$ per cent, just as in our study. However, if we consider a fraction of counterrotating gaseous discs among all extended gaseous discs in S0 galaxies, we see a prominent difference. When S0 galaxies were selected over all types of environment, the fraction of counterrotating gaseous discs was $20-24$ per cent (\citep{bertola92,kuijken_fisher}); more exactly, by combining two samples, \citet{kuijken_fisher} gave $24\pm8$ per cent. In our sample, the fraction of counterrotating gaseous discs is higher, $46\pm14$ per cent. We expected such a trend because \citet{davis_2011} noted a dependence of gas kinematics in the early-type galaxies (mostly S0s in the ATLAS-3D sample) on their environment: dense environment provided tight coincidence between gas and star kinematics while in more sparse environments the fraction of decoupled gaseous kinematics grew. Our isolated S0 galaxies represent an extreme point in this dependency.

Following the logic of \citet{bertola92} according to which when gas is accreted, it should have an arbitrary spin, and so long-slit spectroscopy should reveal an equal fraction of corotating and counterrotating gaseous discs. If one takes into account the possibility for internal gas incidence, the fraction of systems with counterrotating gaseous discs would
be less than 50 per cent. But our results based on the full SCORPIO+SALT sample of galaxies in the strictly sparse environment suggest that the fraction of the gas counterrotations is 50 per cent. Hence, we can conclude that in isolated S0s their gas is virtually always accreted from external sources. But to be certain that the directions of possible gas accretion are distributed isotropically, we must first identify sources of gas accretion. Our galaxies are isolated so they cannot acquire their gas from neighbours of comparable mass/luminosity; the sources of cold gas accretion may be dwarf satellite merging \citep{Kaviraj2009,Kaviraj2011} or perhaps cosmological gas filaments \citep{dekel_flows,keres_flows}. Are they distributed isotropically? This is an open question.

\subsection{Metallicity and origin of gas structures}

Based on diagnostic BPT diagrams we marked out a regions where photoionization by young stars is a dominant excitation mechanism. We binned pure emission spectra in such regions in order to increase signal-to-noise ratio and determined emission line fluxes. We estimated oxygen abundances by utilizing $O2N2$ and $N2$ calibrations proposed by \citet{PPcalib}. The $N2$ calibration was used for bins where signal-to-noise ratios  for \Hb\ and [O\iii] were poor, $S/N<3$. The oxygen abundances as well as the radii for binned regions are shown at Table~\ref{table_abund_bins}.


\begin{table}
{\footnotesize{}
\begin{tabular}{lcc}
\hline\hline
Galaxy & Binning  & 12+$\log$ O/H ([Z/H]$_O$) \\
       & region      &  dex                   \\
\hline
\hline
\multirow{2}{*}{   IC 1608} & (    -51.4;     -31.1) &       8.78 (   0.09)  $\pm   0.47$ \\
                              & (     29.7;      43.4) &       8.80 (   0.11)  $\pm   0.26$ \\
\hline
\multirow{2}{*}{   IC 3152$^{N2}$} & (    -12.6;      -8.1) &       8.73 (   0.04)  $\pm   0.42$ \\
                              & (      4.1;      10.2) &       8.78 (   0.09)  $\pm   0.42$ \\
\hline
   IC 4653 & (     -5.6;       4.6) &       8.57 (  -0.12)  $\pm   0.25$ \\
\hline
\multirow{2}{*}{  NGC 1211$^{N2}$} & (      9.4;      15.8) &       8.72 (   0.03)  $\pm   0.41$ \\
                              & (     32.3;      37.3) &       8.73 (   0.04)  $\pm   0.41$ \\
\hline
\multirow{2}{*}{  NGC 2917} & (    -15.7;      -5.6) &       8.90 (   0.21)  $\pm   0.27$ \\
                              & (      9.6;      19.8) &       8.82 (   0.13)  $\pm   0.26$ \\
\hline
\multirow{2}{*}{  NGC 4240$^{N2}$} & (    -11.6;      -6.6) &       8.80 (   0.11)  $\pm   0.41$ \\
                              & (      4.1;      12.2) &       8.78 (   0.09)  $\pm   0.41$ \\
\hline
\multirow{2}{*}{  NGC 9980$^{N2}$} & (    -13.1;      -4.3) &       8.82 (   0.13)  $\pm   0.42$ \\
                              & (      8.4;      20.8) &       8.71 (   0.02)  $\pm   0.42$ \\
\hline
  NGC 2350 & (     -1.6;       2.0) &       8.68 (  -0.01)  $\pm   0.25$ \\
\hline
\multirow{2}{*}{  NGC 6798$^{N2}$} & (    -34.1;     -27.7) &       8.71 (   0.02)  $\pm   0.41$ \\
                              & (     29.1;      36.6) &       8.73 (   0.04)  $\pm   0.41$ \\
\hline
  NGC 7351 & (     -2.7;       3.8) &       8.64 (  -0.05)  $\pm   0.25$ \\
\hline
\hline
\end{tabular}
}
\caption{Oxygen abundances in star forming regions.}\label{table_abund_bins}
\end{table}

There is a large diversity of galaxy gas/star subsystems which have orbital momentum distinguishing compared to stellar disc: inner polar rings/discs, large-scale outer polar rings/disc, counter-rotating stellar/gaseous discs, polar bulges and inner nuclear discs. Based on dynamic considerations no doubt that these structures are formed from an external material acquired during gravitational interactions with other galaxies or by external accretion processes, which can occur throughout the life of the galaxy. The properties and morphological appearance of resulted decoupled subsystems depend on geometry, angular momentum of accreted material and duration of accretion.

Several acquisition scenario  of the  external gas are considered in the literature:
\begin{enumerate}
\item 
The major dissipative merger. The acquisition of the gas results from a merger of two disc galaxies with unequal mass with orbits close to ``'polar'' \citep{Bekki_1998,Bournaud_2005}.

\item 
Tidal accretion. Decoupled gas structures may form by the disruption of a dwarf companion galaxy orbitating around an early-type system  or by gas stripping from disc galaxy outskirts, captured by galaxy on a parabolic encounter
 \citep{Reshetnikov_Sotnikova_1997,Bournaud_Combes_2003}.

\item
Cold gas accretion from cosmological filaments. Recent theoretical works \citet{keres_flows,dekel_flows,Bournaud_Elmegreen_2009} which are based on numerical simulation, emphasize the important role of cold gas accretion for the formation of disc galaxies. It was shown in context of cosmological simulation, that the cold gas accretion from several filament which are in the different planes, will lead to formation of kinematically  decoupled substructures \citep{Dekel_2009,Roskar_2010,Algorry_2014}. The acquired gas from filament has to possess extremely low metallicity, and taking into account possible mixture of pristine gas with enriched gas stripped from satellites, metallicity shouldn't exceed 1/10 solar value \citep{Agertz_2009}.
\end{enumerate}

The scenario of major mergers is excluded because the isolated lenticular galaxies has no massive ``neighbors'' by definition. The tidal accretion scenario can be considered in the case of interaction with dwarf satellites, excepting tidal gas accretion from outskirts of giant galaxies. The key parameter by means of which it is possible to make a choice between tidal accretion and cold gas accretion from filaments is metallicity of gas. Our measurements of oxygen abundance of ionized gas in the star forming region in isolated S0 galaxies indicates a solar metallicity (see Table~\ref{table_abund_bins}) that excludes accretion from filaments.

\subsection{Luminosity-metallicity relation}

\begin{figure}[ht]
\includegraphics[width=1.0\textwidth]{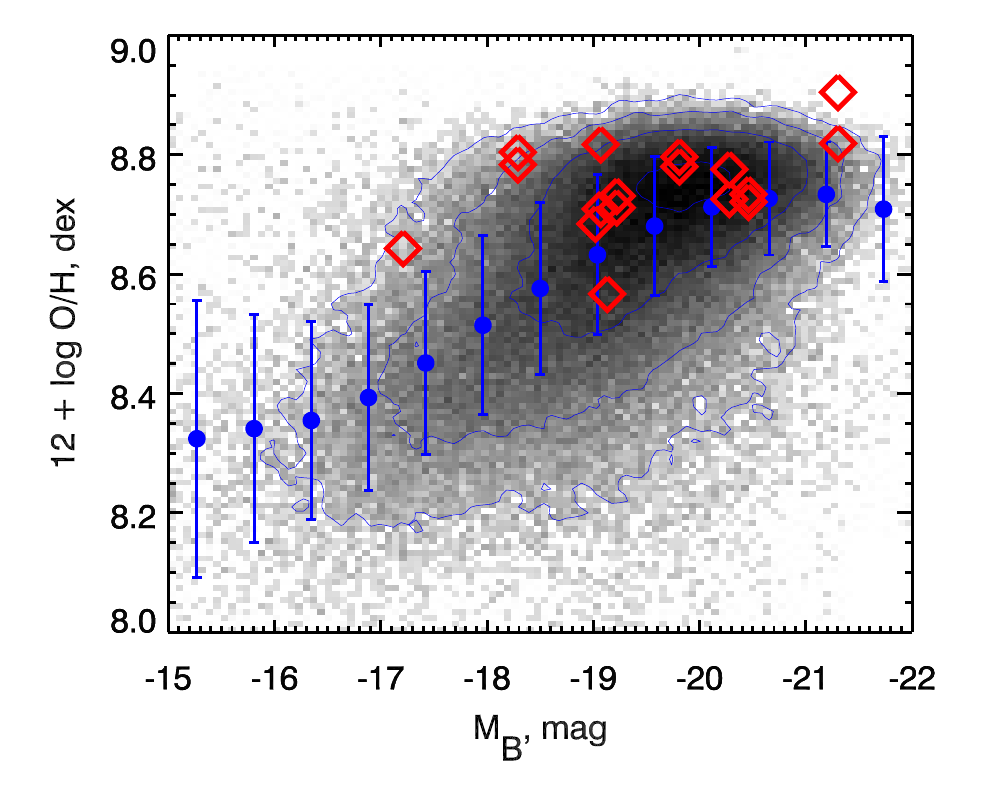}\\
\caption{Dependence gas metallicity on absolute $B$ magnitude. Grey distribution correspond to SDSS data. Blue points shows median values if distribution in bins by magnitude value. Red points shows our measurements in isolated lenticular galaxies.}\label{pics_MZR}
\end{figure} 

The correlation between luminosity or mass of galaxies with the metals enrichment of gas is the well-known observation fact. For the first time this correlation were found for irregular and blue compact galaxies \citep{Lequeux_1979,Kinman_1981}, soon after and for disc galaxies \citep{Rubin_1984}. The subsequent studies of nearby galaxies only strengthened conclusions about mass/luminosity-metallicity relation (see for ex. \citet{Tremonti_2004} and references therein). Various physical mechanisms are proposed for the explanation of mass-metallicity relation: 1) the loss of enriched gas by outflow caused by SN explosions \citep{Tremonti_2004,Kobayashi_2007}; 2) the accretion of pristine gas by inflows from intergalactic medium \cite{Finlator_2008}; 3) variations of the initialmass function with galaxy mass \cite{Koppen_2007}; 4) sformation efficiencies in low-mass galaxies caused by supernova feedback (downsizing) \cite{Brooks_2007} or a combination of them.


At Fig.~\ref{pics_MZR} we plotted luminosity-metallicity for SDSS data and overlapped our measurements of metallicities in star forming regions in isolated S0 galaxies. Despite the wide range of luminosities there is no strong correlation between our measurements of oxygen abundances with absolute $B$ magnitudes. Most of our measurements doesn't match averaged distribution of luminosity-metallicity distribution (blue points). Probably it indicates that chemical evolution of ionized gas in isolated lenticulars doesn't  relate to individual properties of galaxies, what coincides and strengthens a conclusion about an external origin of gas. In the context of scenario of tidal gas accretion from the dwarf satellites, the narrow range of metallicity estimates in our galaxies makes a hint about homogeneous properties of satellites.

\section{Conclusions}
In this work we have presented the results of deep long-slit spectroscopy (total exposures per target 0.3-3 hour) with the 6-m Russian telescope and with Southern African Large Telescope (SALT) for sample of 18 isolated lenticular galaxies. By using full spectral fitting technique as well as Lick indices approach and by analysing of the pure emission line spectra we have obtain that: 
\begin{itemize}
\item
The obtained average ages of the stellar populations in bulges and discs covers a
wide range between 1.5 and $>15$ Gyr, that indicates the absence of distinct epoch of their stellar content formation.
\item
In contrast to galaxies in groups and clusters, the stellar population ages in bulges and discs of isolated
lenticulars tend to be equal, that supports the inefficiency of the bulge rejuvenation in sparse environment. 
\item
Almost all the lenses and rings possess intermediate ages of the stellar populations, within the range of $2-5$ Gyr. On average, the chemical abundances ([Z/H], [Mg/Fe]) in the lenses and rings are the same as in discs, also dynamically the stellar population in lenses and rings do not sufficiently stand out against a discs.
\item
More than half of the sample of isolated lenticulars ($72\pm11$ \%) possess extended emission-line structures; among those, 6 galaxies ($46\pm14$ \%) demonstrate decoupled gas kinematics with respect to their stellar discs. 
\item
We have found starforming off-nuclear regions in 10 galaxies; their gas oxygen abundances are nearly solar that implies tidal gas accretion from gas-rich dwarf satellites rather than accretion from cosmological large-scale structure filaments.

\end{itemize}

%

\renewcommand{\baselinestretch}{0.7}
{\small
\bibliographystyle{apj}
\bibliography{ref_katkov}
}

\end{document}